\documentclass[english,aps,pra]{revtex4}
\usepackage{amsmath}
\usepackage{graphicx}
\usepackage{esint}

\makeatletter

\newcommand{\lyxdot}{.}

\makeatother

\usepackage{babel}
\begin{document}

\title{Transport in time-dependent random potentials}

\author{Yevgeny Krivolapov}

\affiliation{Physics Department, Technion - Israel Institute of Technology, Haifa
32000, Israel.}

\author{Shmuel Fishman}

\affiliation{Physics Department, Technion - Israel Institute of Technology, Haifa
32000, Israel.}
\begin{abstract}
The classical dynamics in stationary potentials that are random both
in space and time is studied. It can be intuitively understood with
the help of Chirikov resonances that are central in the theory of
Chaos, and explored quantitatively in the framework of the Fokker-Planck
equation. In particular, a simple expression for the diffusion coefficient
was obtained in terms of the average power density of the potential.
The resulting anomalous diffusion in velocity is classified into universality
classes. The general theory was applied and numerically tested for
specific examples relevant for optics and atom optics.
\end{abstract}
\maketitle

\section{Introduction}

Potentials that are random both in space in time are models of noise.
Often one seeks for methods of suppressing noise, the present work
on the other-hand is focused on interesting properties of dynamics
generated by noise. It was subject of many sophisticated studies for
nearly 100 years \cite{Langevin1908,Uhlenbeck1930,Sturrock1966,Kampen2007}.
The response to forces resulting of such potentials typically differs
from ordinary diffusion. Specifically, the diffusion coefficients
predicted by such mechanisms sensitively depend on the velocity of
the particles. Also, if the potential is time-dependent, the energies
of the particles will not be constant. The existence of such `anomalous
diffusion' has been demonstrated for classical dynamics with spatially
and temporally fluctuating potentials \cite{Golubovic1991,Rosenbluth1992,Arvedson2006,Bezuglyy2006,Aguer2009,Bezuglyy2012}.
These works claim universal behavior in the sense that for generic
random potentials the diffusion coefficient has a universal power-law
dependence on velocity $v$, such that $D(v)\sim|v|^{-3}$ as $|v|\to\infty$.
This in turn implies that asymptotically in time the average velocity
satisfies $\left\langle v^{2}\right\rangle \sim t^{2/5}$. The average
displacement satisfies $\left\langle x^{2}\right\rangle \sim t^{12/5}$
for one-dimensional systems \cite{Golubovic1991,Rosenbluth1992,Arvedson2006,Bezuglyy2006,Bezuglyy2012}
(faster than ballistic) and $\left\langle x^{2}\right\rangle \sim t^{2}$
(ballistic transport on average) for systems with dimension higher
than one \cite{Rosenbluth1992}. In the present paper we will demonstrate
that this picture should be extended, and we will introduce such an
extension.

The present work is motivated by experiments in optics and in atom
optics. The random potentials induced by optical waves in photonic
lattices \cite{Schwartz2007} and in atom optics experiments \cite{Lye2005,Sanchez-Palencia2007},
rely on transforming an intensity pattern into an effective potential
for the light \cite{Efremidis2002,Fleischer2003} (the former) or
the cold atoms \cite{Lye2005,Sanchez-Palencia2007} (the latter).
Such potentials are naturally described in terms of the Fourier spectrum
of the waves inducing them, and their spectral coefficients are assumed
to be independent random variables. In the regime where the wave-packet
has already acquired large values of velocity, one expects classical
mechanics to provide a reasonable approximation. Therefore, we will
explore here the classical dynamics for potentials, defined by their
spectral content or more precisely their average power spectral density
\cite{Krivolapov2012a}. In particular, we consider the classical
dynamics of a particle in potentials that are random both in space
and in time, emphasizing the spreading of the velocity acquired by
the particle, as time evolves.

In the present paper we will study dynamics of classical particles
in potentials which are defined in terms of their Fourier components.
Two types of potentials will be explored:
\begin{equation}
V^{\left(1\right)}\left(\mathbf{x},t\right)=\int\hat{V}\left(\mathbf{k},\omega\right)\exp i\left(\mathbf{k}\cdot\mathbf{x}-\omega t\right)\mathrm{d}\mathbf{k}\mathrm{d}\omega+c.c.\label{eq:1}
\end{equation}
and
\begin{equation}
V^{\left(2\right)}\left(\mathbf{x},t\right)=\left|V^{\left(1\right)}\left(\mathbf{x},t\right)\right|^{2},\label{eq:2}
\end{equation}
where $\hat{V}\left(\mathbf{k},\omega\right)$ is a random field chosen
such that the distribution of $V^{\left(1\right)}\left(\mathbf{x},t\right)$
is stationary both in time and space. For time dependent potentials
the velocity of the particles may grow, since energy is not conserved.
The dominant mechanism of the growth of the velocity is via Chirikov
resonances, namely, resonances between the particle dynamics and the
external driving. Those resonances occur when the phases in (\ref{eq:1})
are stationary. For a given instantaneous velocity $\mathbf{v}=\dot{\mathbf{x}},$
this happens for
\begin{equation}
\mathbf{k}\cdot\mathbf{v}=\omega,\label{eq:Chirikov_res}
\end{equation}
and a more complicated expression is found for the potential (\ref{eq:2}).
Indeed, transport takes place only in such regions in phase-space
where the density of the resonances is non-zero. The Chirikov resonances
provide an intuitive picture for the understanding of transport in
phase-space. Since, the potentials are given in terms of their Fourier
components it is relatively easy to calculate the spectral content
or average power spectral density (PSD), $S\left(\mathbf{k},\omega\right)$,
(see (\ref{eq:Win-Khin})) and with its help also the diffusion coefficient.
The spread in velocity is then calculated using the Fokker-Planck
(FP) approximation, assuming that the decay of the potential correlations
is sufficiently rapid. The random potentials of the form (\ref{eq:1})
and (\ref{eq:2}) are natural in optics realizations since these are
formed by a superposition of light beams. In recent experiments in
optics \cite{Schwartz2007,Levi2011}, light propagates paraxially
in a disordered potential, which is produced by utilizing the photo-sensitivity
of the medium \cite{Efremidis2002,Fleischer2003}. In the paraxial
approximation a probe beam satisfies a Schrödinger like equation where
the refractive index plays the role of the potential, the propagation
direction plays the role of time and the dispersion relation is approximately
$\omega\left(k\right)=k^{2}/2$. Since the index of refraction is
proportional to intensity of the electric field of the writing beam,
that varies like $V^{\left(1\right)}\left(\mathbf{x},t\right)$ of
(\ref{eq:1}), the resulting potential will have the form of $V^{\left(2\right)}\left(\mathbf{x},t\right)$
(\ref{eq:2}). Potentials of this type are also relevant for atom
optics. There the electric field induced a dipole moment in a neutral
atom which interacts with the electric field and results in a potential
which is proportional to the intensity of the electric field. The
time dependence is induced by a modulation of the intensity of the
electric field, such that it is independent of $\mathbf{k}$.

Although this work was motivated by experiments in optics, where the
dynamics is wave dynamics, the classical dynamics of particles in
such potentials is a fundamental question by itself. Therefore, in
this work we will study the classical dynamics of particles in potentials
of the form (\ref{eq:1}) and (\ref{eq:2}). Nevertheless, since it
is generally believed that for large velocity (or short wave length)
wave dynamics is approximated by classical dynamics, we expect at
least qualitatively to describe the original wave problem.

We will classify the various systems into universality classes according
to the velocity dependence of the diffusion coefficient in large and
small velocity limits. The regimes of the validity of the FP approximation
will be studied as well, and the crossover between uniform acceleration
and diffusive behavior (in velocity) will be studied. The formalism
will be demonstrated for specific examples of the form (\ref{eq:1})
and (\ref{eq:2}). Our main result is that although the universality
of the form $D\left(v\right)\sim v^{-3}$ holds always for dimensions
higher than one, for one dimensional systems there is a large variety
of novel universality classes. A classification scheme for such classes
will be presented and it will be demonstrated how these can be realized.

In Sec. \ref{sec:Time-domains} the time required for the particles
to `discover' the fluctuations of the potential will be evaluated.
After this time we expect the dynamics to be controlled by the FP
equation. An expression for the diffusion coefficient in terms of
the spectral content or more precisely the average power spectral
density (PSD) is derived in Sec. \ref{sec:Anomalous-diffusion}, it
is computed explicitly and its significance is discussed for specific
examples. A scaling property of the FP equation is presented in Sec.
\ref{sec:Scaling-properties}. A detailed calculation of the diffusion
coefficient in velocity for dimensions two and higher is presented
in Sec. \ref{sec:Potentials-sperposition} for potentials of the form
of (\ref{eq:1}), and in Sec. \ref{sec:Potentials-Optics} for potentials
of the form (\ref{eq:2}). In Sec. \ref{sec:Anomalous-diffusion-1D}
the rich variety of the universality classes found for one-dimensional
systems is presented and the asymptotic expansion in powers of $v^{-1}$
is discussed. The results are summarized and discussed in Sec. \ref{sec:Discussion}.

\section{\label{sec:Time-domains}Time domains}

In this section we discuss the typical time scale to observe the anomalous
diffusion. For this purpose we assume an initial distribution that
is narrow in phase-space. Initially the velocity is growing linearly
in time (acceleration), since it takes time for the fluctuations of
the potential to become effective. First we define the characteristic
length and time scales of the potential. Consider the potential given
in (\ref{eq:1}), let $l_{x}$ be the length over which the potential
has an almost constant derivative, resulting in a constant force.
It satisfies,
\begin{equation}
l_{x}=\sqrt{\frac{\left\langle V^{2}\right\rangle -\left\langle V\right\rangle ^{2}}{\left\langle \left|\nabla V\right|^{2}\right\rangle }}.
\end{equation}
In a similar way we define the characteristic time scale $l_{t}$,
\begin{equation}
l_{t}=\sqrt{\frac{\left\langle V^{2}\right\rangle -\left\langle V\right\rangle ^{2}}{\left\langle \left(\frac{\partial V}{\partial t}\right)^{2}\right\rangle }.}
\end{equation}
Using,
\begin{eqnarray}
\left\langle V^{2}\right\rangle  & = & \int\mathrm{d}\omega_{1}\mathrm{d}\omega_{2}\int\mathrm{d}\mathbf{k}_{1}\mathrm{d}\mathbf{k}_{2}\left\langle \hat{V}\left(\mathbf{k}_{1},\omega_{2}\right)\hat{V}^{*}\left(\mathbf{k}_{2},\omega_{2}\right)\right\rangle \exp i\left(\left(\mathbf{k}_{1}-\mathbf{k}_{2}\right)\cdot\mathbf{x}-\left(\omega_{1}-\omega_{2}\right)t\right),\nonumber \\
 & = & \int\mathrm{d}\omega\int\mathrm{d}\mathbf{k}\, S\left(\mathbf{k},\omega\right),
\end{eqnarray}
one finds,
\begin{eqnarray}
\left\langle \left|\nabla V\right|^{2}\right\rangle  & = & \int\mathrm{d}\omega_{1}\mathrm{d}\omega_{2}\int\mathrm{d}\mathbf{k}_{1}\mathrm{d}\mathbf{k}_{2}\mathbf{k}_{1}\cdot\mathbf{k}_{2}\left\langle \hat{V}\left(\mathbf{k}_{1},\omega_{2}\right)\hat{V}^{*}\left(\mathbf{k}_{2},\omega_{2}\right)\right\rangle \exp i\left(\left(\mathbf{k}_{1}-\mathbf{k}_{2}\right)\cdot\mathbf{x}-\left(\omega_{1}-\omega_{2}\right)t\right),\nonumber \\
 & = & \int\mathrm{d}\omega\int\mathrm{d}\mathbf{k}\, k^{2}S\left(\mathbf{k},\omega\right),
\end{eqnarray}
where $S\left(\mathbf{k},\omega\right)$ is the average power spectral
density (PSD) defined in (\ref{eq:Weiner-Khintchin}). Therefore the
characteristic length scale, $l_{x},$ is given by,
\begin{equation}
l_{x}^{2}=\frac{\int\mathrm{d}\omega\int\mathrm{d}\mathbf{k}\, S\left(\mathbf{k},\omega\right)}{\int\mathrm{d}\omega\int\mathrm{d}\mathbf{k}\, k^{2}S\left(\mathbf{k},\omega\right)},\label{eq:corr_def}
\end{equation}
and similarly the characteristic time scale, $l_{t}$, is given by,
\begin{equation}
l_{t}^{2}=\frac{\int\mathrm{d}\omega\int\mathrm{d}\mathbf{k}\, S\left(\mathbf{k},\omega\right)}{\int\mathrm{d}\omega\int\mathrm{d}\mathbf{k}\,\omega^{2}S\left(\mathbf{k},\omega\right)}.
\end{equation}
Note, that $l_{t}$ and $l_{x}$ are different from the exponential
decay rates of correlations. The difference follows from the fact
that we are interested in the typical lengths in both time and space
where the force is constant, and not in the asymptotic behavior of
the correlation function. For times $t<l_{t}$ the force resulting
from a potential is almost constant, provided that the particle is
not displaced more than $l_{x}$. For a weak force,
\begin{equation}
F_{0}<\frac{l_{x}}{l_{t}^{2}},\label{eq:F_cond}
\end{equation}
the particle will travel a distance less than $l_{x}$ in time $l_{t}$,
therefore the cross-over out of the uniform acceleration regime takes
place at time,
\begin{equation}
t_{*}\sim l_{t}.\label{eq:t_star}
\end{equation}
If the force is strong so that (\ref{eq:F_cond}) is not satisfied,
the particle will reach ,$l_{x}$ at a time shorter than $l_{t}$
and therefore $t_{*}$ will be,
\begin{equation}
t_{*}\sim\sqrt{\frac{l_{x}}{F_{0}}}.
\end{equation}
In this work we will consider initial conditions of particles with
a narrow distribution both in velocity and position, such that $\delta v<l_{x}/l_{t}$,
where $\delta v$ is the width of the distribution centered at $v=0$.
For these initial conditions uniform acceleration will take place
for time $t<l_{t}$, while for longer time-scales the velocity will
exhibit anomalous diffusion. In this work we will assume that the
force is weak, such that (\ref{eq:F_cond}) holds and the crossover
to the diffusive regime will take place at (\ref{eq:t_star}).

\section{Anomalous diffusion in velocity\label{sec:Anomalous-diffusion}}

In this section the variation of the velocity of particles by potentials
$V\left(\mathbf{x},t\right)$, random both in space and time will
be studied on time scales longer than $t_{*}$. The equations of motion
are,
\begin{eqnarray}
\frac{\mathrm{d}\mathbf{v}}{\mathrm{d}t} & = & -\nabla V\left(\mathbf{x},t\right)\label{eq:Hamilton_eq}\\
\frac{\mathrm{d}\mathbf{x}}{\mathrm{d}t} & = & \mathbf{v},\nonumber
\end{eqnarray}
where $\mathbf{v}$ and $\mathbf{x}$ are the instantaneous velocity
and position. For simplicity we set the mass to be one. We will assume
a stationary potential both in space and time and an isotropic distribution
of the potential. Therefore, its correlation function is
\begin{equation}
\left\langle V\left(\mathbf{x}_{1},t_{1}\right)V\left(\mathbf{x}_{2},t_{2}\right)\right\rangle =C\left(\mathbf{x}_{1}-\mathbf{x}_{2},t_{1}-t_{2}\right),
\end{equation}
and we may take the average of the potentials, that is a constant,
to vanish, namely,
\begin{equation}
\left\langle V\left(\mathbf{x},t\right)\right\rangle =0.
\end{equation}
The angular brackets $\left\langle .\right\rangle $ denote the average
over the ensemble of random potentials. The Weiner-Khinchin theorem
implies,
\begin{equation}
C\left(\mathbf{x},t\right)=\int\mathrm{d}\omega\int\mathrm{d}\mathbf{k}\, S\left(\mathbf{k},\omega\right)\exp i\left(\mathbf{k}\cdot\mathbf{x}-\omega t\right),\label{eq:Weiner-Khintchin}
\end{equation}
where $S\left(\mathbf{k},\omega\right)$ is the average power spectral
density (PSD) of the potential, that is its spectral content.

The Fokker-Planck (FP) equation for the velocity is given by \cite{Kampen2007},
\begin{equation}
\frac{\partial P}{\partial t}=\left(\frac{\partial}{\partial v_{i}}D_{ij}\frac{\partial}{\partial v_{j}}\right)P,\label{eq:FP_momentum}
\end{equation}
where $P\left(\mathbf{v},t\right)$ is the probability density and
$D_{ij}$ is the diffusion tensor, given by the expression,
\begin{equation}
D_{ij}\left(v\right)=\frac{1}{2}\int_{-\infty}^{\infty}\left\langle F_{i}\left(\mathbf{v}\tau,\tau\right)F_{j}\left(0,0\right)\right\rangle \mathrm{d}\tau,
\end{equation}
where
\begin{equation}
\mathbf{F}=-\nabla V
\end{equation}
is the force. Here one assumes that the correlations decay sufficiently
fast and the force is sufficiently weak, so that the variation of
the velocity on the time-scale of decay of correlations is negligible,
and
\begin{equation}
\mathbf{x_{1}-x_{2}}=\mathbf{v}\left(t_{1}-t_{2}\right),\label{eq:FP_approx}
\end{equation}
this is the usual approximation used to derive the Fokker-Planck equation
\cite{Kampen2007}. Here and in the rest of the paper summation over
repeated indexes is assumed. We will now calculate the force correlation
function,
\begin{equation}
K_{ij}\left(\mathbf{x}_{1}-\mathbf{x}_{2},t_{1}-t_{2}\right)=\left\langle F_{i}\left(\mathbf{x}_{1},t_{1}\right)F_{j}\left(\mathbf{x}_{2},t_{2}\right)\right\rangle ,
\end{equation}
which can be written using the potential correlation function,
\begin{equation}
K_{ij}\left(\mathbf{x}_{1}-\mathbf{x}_{2},t_{1}-t_{2}\right)=\frac{\partial^{2}}{\partial x_{1,i}\partial x_{2,j}}\left\langle V\left(\mathbf{x}_{1},t_{1}\right)V\left(\mathbf{x}_{2},t_{2}\right)\right\rangle =-\frac{\partial^{2}C\left(\mathbf{x},t\right)}{\partial x_{i}\partial x_{j}}|_{\mathbf{x}=\mathbf{x}_{1}-\mathbf{x}_{2}}.
\end{equation}
Using (\ref{eq:Weiner-Khintchin}) one finds,
\begin{equation}
K_{ij}\left(\mathbf{x}_{1}-\mathbf{x}_{2},t_{1}-t_{2}\right)=\int\mathrm{d}\omega\int\mathrm{d}\mathbf{k}\, k_{i}k_{j}S\left(\mathbf{k},\omega\right)\exp i\left[\mathbf{k}\cdot\left(\mathbf{x_{1}-x_{2}}\right)-\omega\left(t_{1}-t_{2}\right)\right],
\end{equation}
and the diffusion tensor is,
\begin{equation}
D_{ij}\left(\mathbf{v}\right)=\frac{1}{2}\int_{-\infty}^{\infty}K_{ij}\left(\mathbf{v}\tau,\tau\right)\mathrm{d}\tau.
\end{equation}
Which can be written as,
\begin{eqnarray}
D_{ij}\left(\mathbf{v}\right) & = & \frac{1}{2}\int_{-\infty}^{\infty}\mathrm{d}\tau\int\mathrm{d}\omega\int\mathrm{d}\mathbf{k}\, k_{i}k_{j}S\left(\mathbf{k},\omega\right)\exp i\tau\left(\mathbf{k}\cdot\mathbf{v}-\omega\right).
\end{eqnarray}
Taking the integral over $\tau$ first, gives a delta function
\begin{eqnarray}
D_{ij}\left(\mathbf{v}\right) & = & \pi\int\mathrm{d}\omega\int\mathrm{d}\mathbf{k}\, k_{i}k_{j}S\left(\mathbf{k},\omega\right)\delta\left(\mathbf{k}\cdot\mathbf{v}-\omega\right).\label{eq:Dij_delta}
\end{eqnarray}
Then the integral over $\omega$ can be easily performed to give,
\begin{equation}
D_{ij}\left(\mathbf{v}\right)=\pi\int\mathrm{d}\mathbf{k}\, k_{i}k_{j}S\left(\mathbf{k},\mathbf{k}\cdot\mathbf{v}\right).
\end{equation}
Since $S\left(\mathbf{k},\omega\right)$ is integrable by definition,
we will define a function $F\left(k,\omega\right)$, so that
\begin{equation}
S\left(\mathbf{k},\omega\right)=\frac{\partial^{2}}{\partial\omega^{2}}F\left(\mathbf{k},\omega\right).\label{eq:F}
\end{equation}
Then, the diffusion tensor can be written as,
\begin{equation}
D_{ij}\left(\mathbf{v}\right)=\frac{\partial^{2}}{\partial v_{i}\partial v_{j}}I\left(v\right),\label{eq:Dij}
\end{equation}
where
\begin{equation}
I\left(v\right)=\pi\int\mathrm{d}\mathbf{k}\, F\left(\mathbf{k},\mathbf{k}\cdot\mathbf{v}\right),\label{eq:I_v}
\end{equation}
and we have used the fact that the integral is isotropic. Equation
(\ref{eq:Dij}) can be decomposed to,
\begin{eqnarray}
D_{ij}\left(\mathbf{v}\right) & = & \frac{\partial^{2}}{\partial v_{i}\partial v_{j}}I\left(v\right)=\frac{\partial}{\partial v_{i}}\left(\frac{v_{j}}{v}\frac{\partial I}{\partial v}\right)\label{eq:Dij_I}\\
 & = & \left(\frac{\delta_{ij}v-v_{i}v_{j}/v}{v^{2}}\right)\frac{\partial I}{\partial v}+\frac{v_{i}v_{j}}{v^{2}}\frac{\partial^{2}I}{\partial v^{2}}.\nonumber
\end{eqnarray}
For a general isotropic diffusion tensor of the form,
\begin{equation}
D_{ij}\left(\mathbf{v}\right)=\left(\delta_{ij}-\frac{v_{i}v_{j}}{v^{2}}\right)f_{T}\left(v\right)+\frac{v_{i}v_{j}}{v^{2}}f_{P}\left(v\right),\label{eq:General_isotropic_tensor}
\end{equation}
the operator on the RHS of (\ref{eq:FP_momentum}) is,
\begin{equation}
\hat{L}_{0}=\frac{\partial}{\partial v_{i}}D_{ij}\left(\mathbf{v}\right)\frac{\partial}{\partial v_{j}}=v^{-\left(d-1\right)}\frac{\partial}{\partial v}v^{\left(d-1\right)}f_{P}\left(v\right)\frac{\partial}{\partial v}.\label{eq:L0}
\end{equation}
We turn now to justify this result. It is convenient to use hyper-spherical
variables, such that,
\begin{equation}
\frac{\partial}{\partial v_{i}}=\frac{\partial v}{\partial v_{i}}\frac{\partial}{\partial v}+\sum_{j=1}^{d-1}\frac{\partial\phi_{j}}{\partial v_{i}}\frac{\partial}{\partial\phi_{j}}=\frac{v_{i}}{v}\frac{\partial}{\partial v}+\sum_{j=1}^{d-1}\frac{\partial\phi_{j}}{\partial v_{i}}\frac{\partial}{\partial\phi_{j}},
\end{equation}
where $d$ is the dimension of space and $\phi_{j}$ are the angle
coordinates. Using (\ref{eq:General_isotropic_tensor}) and the new
variables the operator (\ref{eq:L0}) reduces to,
\begin{eqnarray}
\hat{L}_{0} & = & \frac{\partial}{\partial v_{i}}\left[\left(\delta_{ij}-\frac{v_{i}v_{j}}{v^{2}}\right)f_{T}\left(v\right)+\frac{v_{i}v_{j}}{v^{2}}f_{P}\left(v\right)\right]\frac{\partial}{\partial v_{j}}\nonumber \\
 & = & \frac{\partial}{\partial v_{i}}\left[\left(\delta_{ij}-\frac{v_{i}v_{j}}{v^{2}}\right)f_{T}\left(v\right)+\frac{v_{i}v_{j}}{v^{2}}f_{P}\left(v\right)\right]\frac{v_{j}}{v}\frac{\partial}{\partial v}\nonumber \\
 & = & \frac{\partial}{\partial v_{i}}\left[\left(v_{i}-v_{i}\right)\frac{1}{v}f_{T}\left(v\right)+\frac{v_{i}}{v}f_{P}\left(v\right)\right]\frac{\partial}{\partial v}\nonumber \\
 & = & \frac{\partial}{\partial v_{i}}\frac{v_{i}}{v}f_{P}\left(v\right)\frac{\partial}{\partial v}.
\end{eqnarray}
Which using the chain rule can be written as,
\begin{eqnarray}
\hat{L}_{0} & = & \left(\frac{\delta_{ii}v-v_{i}v_{i}/v}{v^{2}}\right)f_{P}\left(v\right)\frac{\partial}{\partial v}+\frac{\partial}{\partial v}f_{P}\left(v\right)\frac{\partial}{\partial v}\nonumber \\
 & = & \frac{1}{v}\left(d-1\right)f_{P}\left(v\right)\frac{\partial}{\partial v}+\frac{\partial}{\partial v}f_{P}\left(v\right)\frac{\partial}{\partial v}\nonumber \\
 & = & v^{-\left(d-1\right)}\frac{\partial}{\partial v}v^{\left(d-1\right)}f_{P}\left(v\right)\frac{\partial}{\partial v}.
\end{eqnarray}
Therefore, combining (\ref{eq:Dij_I}) and (\ref{eq:L0}) one identifies
$f_{P}\left(v\right)=\partial^{2}I/\partial v^{2}$. Therefore, the
operator $\hat{L}_{0}$ is given by,
\begin{equation}
\hat{L}_{0}=v^{-\left(d-1\right)}\frac{\partial}{\partial v}v^{\left(d-1\right)}D\left(v\right)\frac{\partial}{\partial v}.
\end{equation}
The diffusion coefficient $D\left(v\right)$ is found to be,
\begin{equation}
D\left(v\right)=\frac{\partial^{2}I}{\partial v^{2}}.
\end{equation}
It can be simplified using the definition (\ref{eq:I_v}) and (\ref{eq:F}),
\begin{equation}
D\left(v\right)=\pi\int\mathrm{d}\mathbf{k}\,\left(\mathbf{k}\cdot\hat{v}\right)^{2}S\left(\mathbf{k},\mathbf{k}\cdot\mathbf{v}\right).\label{eq:Dv_d}
\end{equation}
Using (\ref{eq:Dv_d}) and aligning the $x-$component of $\mathbf{k}$
with velocity we get,
\begin{equation}
D\left(v\right)=\frac{S_{d}}{2}\int_{0}^{2\pi}\mathrm{d}\theta\int_{0}^{\infty}\mathrm{d}k\, k^{d+1}\cos^{2}\theta S\left(k,kv\cos\theta\right),\label{eq:Dv_angles}
\end{equation}
where $S_{d}$ is the surface of the $d-$dimensional hyper-sphere.
Changing variables to, $y=v\cos\theta$ we get
\begin{equation}
D\left(v\right)=\frac{S_{d}}{v^{3}}\int_{-v}^{v}\mathrm{d}y\frac{y^{2}}{\sqrt{1-\left(y/v\right)^{2}}}\int_{0}^{\infty}\mathrm{d}k\, k^{d+1}S\left(k,ky\right),\label{eq:Dv_no_deltas}
\end{equation}
where the factor of two was eliminated due the multiplicity. This
expression assumes implicitly that $d>1$. For one dimensional systems
there is no integration over the angle (see Eq. (\ref{eq:Dv_1d})
in Sec. \ref{sec:Anomalous-diffusion-1D}). The different dependence
on the velocity of (\ref{eq:Dv_no_deltas}) and (\ref{eq:Dv_1d})
is the origin of the richness of the one dimensional behavior. For
large velocities (\ref{eq:Dv_no_deltas}) has the following asymptotic
behavior,
\begin{equation}
D\left(v\right)\sim\frac{D_{3}}{v^{3}},\label{eq:D_univ}
\end{equation}
with
\begin{equation}
D_{3}=2S_{d}\int_{0}^{\infty}\mathrm{d}y\int_{0}^{\infty}\mathrm{d}k\, y^{2}k^{d+1}S\left(k,ky\right).\label{eq:D0_univ}
\end{equation}
Therefore, for dimensions higher than one the asymptotic behavior
of the diffusion coefficient is indeed universal for any choice of
the PSD and is given by (\ref{eq:D_univ}). The diffusion coefficient
for zero velocity using (\ref{eq:Dv_angles}) is given by,
\begin{equation}
D\left(0\right)=\frac{S_{d}}{4}\int_{0}^{\infty}\mathrm{d}k\, k^{d+1}S\left(k,0\right).\label{eq:D_zero}
\end{equation}
If $D_{3}\neq0$, as is the case for $d\geq2$, the Fokker-Planck
equation for the velocity (\ref{eq:FP_momentum}) is asymptotically
equivalent to the equation,
\begin{equation}
\frac{\partial P}{\partial t}=\left(v^{-\left(d-1\right)}\frac{\partial}{\partial v}v^{\left(d-1\right)}\frac{D_{3}}{v^{3}}\frac{\partial}{\partial v}\right)P,\label{eq:FP_mom_asymp}
\end{equation}
which has the scaling solution
\begin{equation}
P\left(v,t\right)=\frac{1}{t^{d/5}}g\left(\frac{v^{5}}{t}\right).
\end{equation}
The resulting growth of the mean kinetic energy is,
\begin{equation}
\frac{1}{2}\left\langle v^{2}\right\rangle \sim t^{2/5}.\label{eq:generic_coeff}
\end{equation}
This behavior is considered in the literature as universal \cite{Golubovic1991,Arvedson2006,Bezuglyy2006,Bezuglyy2012}.
For dimensions two and higher this is indeed the case, however for
one dimensional systems other behaviors are possible, as will be demonstrated
in Sec. \ref{sec:Anomalous-diffusion-1D}.

\section{\label{sec:Scaling-properties}Scaling properties}

Let us assume that the average power spectral density, $S\left(\mathbf{k},\omega\right)$,
has natural spatial and temporal frequency scales, $k_{0}$ and $\omega_{0}$,
such that,
\begin{equation}
S\left(\mathbf{k},\omega\right)=\frac{V_{0}^{2}}{k_{0}^{d}\omega_{0}}\tilde{S}\left(\frac{\mathbf{k}}{k_{0}},\frac{\omega}{\omega_{0}}\right),
\end{equation}
where $V_{0}$ is a constant which determines the strength of the
potential. Then (\ref{eq:Dv_d}) takes the form,
\begin{equation}
D\left(v\right)=\pi\frac{V_{0}^{2}}{k_{0}^{d}\omega_{0}}\int\left(\mathbf{k}\cdot\hat{v}\right)^{2}\tilde{S}\left(\frac{\mathbf{k}}{k_{0}},\frac{\mathbf{k}\cdot\mathbf{v}}{\omega_{0}}\right)\mathrm{d}\mathbf{k},
\end{equation}
where $\widetilde{S}$ is a dimensionless PSD. Rescaling the variables
\begin{equation}
\mathbf{k}=\mathbf{k}'k_{0}\qquad\mathbf{v}=\mathbf{v}'v_{0}\qquad v_{0}\equiv\frac{\omega_{0}}{k_{0}},
\end{equation}
gives
\begin{equation}
D\left(v'\right)=\pi V_{0}^{2}\frac{k_{0}^{2}}{\omega_{0}}\int\left(\mathbf{k}'\cdot\hat{v}\right)^{2}\tilde{S}\left(\mathbf{k}',\mathbf{k}'\cdot\mathbf{v}'\right)\mathrm{d}\mathbf{k}'.
\end{equation}
The Fokker-Planck equation for the velocity (\ref{eq:FP_momentum})
is invariant under the transformation of variables,
\begin{eqnarray}
\mathbf{v} & \to & \mathbf{v}'\nonumber \\
t & \to & t'\left(\pi I_{0}^{2}\frac{k_{0}^{2}}{\omega_{0}}\right)^{-1}.
\end{eqnarray}

\section{\label{sec:Special-potentials}Special potentials}

In this section the spreading in phase-space is studied for specific
potentials. Potentials of the form (\ref{eq:1}) will be studied in
subsection \ref{sec:Potentials-sperposition} while potentials of
the form (\ref{eq:2}) will be studied in subsection \ref{sec:Potentials-Optics}.
In this section the emphasis will be on the behavior for dimensions
higher than one, while one dimensional systems will be analyzed in
the next section.

\subsection{\label{sec:Potentials-sperposition}Potentials which are a superposition
of waves}

Natural potentials to consider are potentials which are composed of
a superposition of standing waves (c.f., Eq. (\ref{eq:1})),
\begin{equation}
V\left(\mathbf{x},t\right)=\int\hat{V}\left(\mathbf{k}\right)\exp i\left(\mathbf{k}\cdot\mathbf{x}-\omega\left(\mathbf{k}\right)t\right)\mathrm{d}\mathbf{k}+c.c.\label{eq:Cos_potential}
\end{equation}
with a dispersion relation $\omega\left(\mathbf{k}\right)$. We will
assume that the amplitudes and the wave numbers are independent random
variables. Leading to,

\begin{equation}
\left\langle \hat{V}\left(\mathbf{k}_{1}\right)\hat{V}^{*}\left(\mathbf{k}_{2}\right)\right\rangle =V_{0}^{2}f\left(\mathbf{k}_{1}\right)\delta\left(\mathbf{k}_{1}-\mathbf{k}_{2}\right),
\end{equation}
where $\left\langle \left|\hat{V}\left(\mathbf{k}\right)\right|^{2}\right\rangle =V_{0}^{2}$
and $f\left(\mathbf{k}\right)$ is the distribution of the wave numbers.
The correlation function of the potential is given by,
\begin{equation}
C\left(\mathbf{x}_{1}-\mathbf{x}_{2},t_{1}-t_{2}\right)=V_{0}^{2}\int f\left(\mathbf{k}\right)\exp i\left(\mathbf{k}\cdot\left(\mathbf{x}_{1}-\mathbf{x}_{2}\right)-\omega\left(\mathbf{k}\right)\left(t_{1}-t_{2}\right)\right)\mathrm{d}\mathbf{k}+c.c.
\end{equation}
Using (\ref{eq:Weiner-Khintchin}) the PSD is,

\begin{eqnarray}
S\left(\mathbf{q},\omega\right) & = & \frac{1}{\left(2\pi\right)^{d+1}}\int C\left(\mathbf{x},t\right)\exp-i\left(\mathbf{q}\cdot\mathbf{x}-\omega t\right)\mathrm{d}\mathbf{x}\mathrm{d}t\\
 & = & \frac{1}{\left(2\pi\right)^{d+1}}V_{0}^{2}\int f\left(\mathbf{k}\right)\exp i\left(\left(\mathbf{k}-\mathbf{q}\right)\cdot\mathbf{x}-\left(\omega\left(\mathbf{k}\right)-\omega\right)t\right)\mathrm{d}\mathbf{k}\mathrm{d}\mathbf{x}\mathrm{d}t\nonumber \\
 & + & \frac{1}{\left(2\pi\right)^{d+1}}V_{0}^{2}\int f\left(\mathbf{k}\right)\exp-i\left(\left(\mathbf{k}+\mathbf{q}\right)\cdot\mathbf{x}-\left(\omega\left(\mathbf{k}\right)+\omega\right)t\right)\mathrm{d}\mathbf{k}\mathrm{d}\mathbf{x}\mathrm{d}t.\nonumber
\end{eqnarray}
By making the integrals over $\mathbf{x}$ and $t$ we get,
\begin{eqnarray}
S\left(\mathbf{q},\omega\right) & = & V_{0}^{2}f\left(\mathbf{q}\right)\delta\left(\omega-\omega\left(\mathbf{q}\right)\right)+V_{0}^{2}f\left(-\mathbf{q}\right)\delta\left(\omega+\omega\left(\mathbf{q}\right)\right).\label{eq:PSD_disspersion}
\end{eqnarray}
Starting from (\ref{eq:Dv_no_deltas}) and assuming that both $f\left(\mathbf{k}\right)$
and $\omega\left(\mathbf{k}\right)$ are isotropic gives,
\begin{equation}
D\left(v\right)=\frac{S_{d}V_{0}^{2}}{v^{3}}\int_{-v}^{v}\mathrm{d}y\frac{y^{2}}{\sqrt{1-\left(y/v\right)^{2}}}\int_{0}^{\infty}\mathrm{d}k\, k^{d+1}f\left(k\right)\left[\delta\left(ky-\omega\left(k\right)\right)+\delta\left(ky+\omega\left(k\right)\right)\right].
\end{equation}
We can now perform the integral over $y$ and obtain,
\begin{equation}
D\left(v\right)=\frac{2S_{d}}{v^{3}}V_{0}^{2}\int_{0}^{\infty}\mathrm{d}k\,\frac{k^{d-2}f\left(k\right)\omega^{2}\left(k\right)}{\sqrt{1-\omega^{2}\left(k\right)/\left(v^{2}k^{2}\right)}}H\left(v-\frac{\omega\left(k\right)}{k}\right),\label{eq:Dv_disspersion}
\end{equation}
where $H\left(.\right)$ is a Heaviside step function, which follows
from the restriction of the delta function. In the limit of large
velocities we have,
\begin{equation}
D\left(v\right)\sim\frac{D_{3}}{v^{3}},
\end{equation}
with
\begin{equation}
D_{3}=2S_{d}V_{0}^{2}\int_{0}^{\infty}\mathrm{d}k\, k^{d-2}f\left(k\right)\omega^{2}\left(k\right).
\end{equation}
For zero velocity using (\ref{eq:D_zero}) we have,
\begin{eqnarray}
D\left(0\right) & = & \frac{S_{d}}{2}V_{0}^{2}\int_{0}^{\infty}\mathrm{d}k\, k^{d+1}f\left(k\right)\delta\left(\omega\left(k\right)\right).\\
 & = & \frac{S_{d}}{2}V_{0}^{2}\int_{0}^{\infty}\mathrm{d}k\, k^{d+1}\frac{f\left(k\right)}{\left|\omega'\left(0\right)\right|}\delta\left(k\right)=0,\nonumber
\end{eqnarray}
where we have assumed that
\begin{equation}
\omega\left(k\right)=0,
\end{equation}
has only one solution $k=0$, which is a general property of physically
relevant dispersion relations.

\begin{figure}
\begin{centering}
\includegraphics{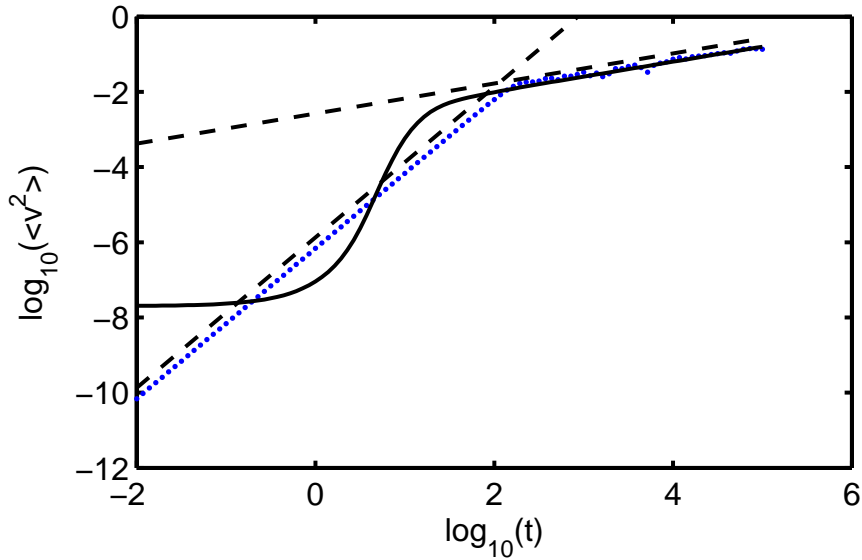}
\par\end{centering}

\caption{\label{fig:2d-cos-disk}A log-log plot of average squared velocity
as a function of time for a two dimensional system with the potential
(\ref{eq:Cos_potential}) and a distribution of wave-numbers (\ref{eq:fk_uniform_disk}).
The blue dots represent the result of the Monte-Carlo solution of
(\ref{eq:Hamilton_eq}) averaged over $20$ realizations and the black
solid line is the numerical solution of the Fokker-Planck equation
for the velocity (\ref{eq:FP_momentum}). The dashed black and red
lines are guides for the eye with the corresponding slopes of 2 and
2/5. The initial condition was a narrow distribution of velocities
for the Fokker-Planck and $\mathbf{x}=\mathbf{v}=0$ for the Monte-Carlo
calculation. The parameters used for this simulation are, $V_{0}=10^{-2}$,
$k_{R}=0.1$.}
\end{figure}

We will proceed by calculating the diffusion coefficient for a particular
choice of the dispersion relation, $\omega\left(k\right)=k^{2}/2$,
and a uniform density of wave numbers,
\begin{equation}
f\left(k\right)=\begin{cases}
\frac{1}{V_{d}k_{R}^{d}} & 0<k<k_{R}\\
0 & k>k_{R}
\end{cases},\label{eq:fk_uniform_disk}
\end{equation}
where $V_{d}$ is the volume of the $d-$ dimensional unit hyper sphere.
Therefore, we have
\begin{equation}
D\left(v\right)=\frac{dV_{0}^{2}}{2k_{R}^{d}v^{3}}\begin{cases}
\int_{0}^{2v}\mathrm{d}k\,\frac{k^{d+2}}{\sqrt{1-k^{2}/\left(4v^{2}\right)}} & v<k_{R}/2\\
\int_{0}^{k_{R}}\mathrm{d}k\,\frac{k^{d+2}}{\sqrt{1-k^{2}/\left(4v^{2}\right)}} & v>k_{R}/2
\end{cases},
\end{equation}
where we have used the fact $S_{d}/V_{d}=d$. Taking the integrals
gives,
\begin{equation}
D\left(v\right)=\frac{dV_{0}^{2}}{2k_{R}^{d}v^{3}}\begin{cases}
\frac{2^{d+2}\sqrt{\pi}\Gamma\left(\frac{d+3}{2}\right)}{\Gamma\left(2+\frac{d}{2}\right)}v^{d+3} & v<k_{R}/2\\
\frac{k_{R}^{d+3}}{d+3}\,_{1}F_{2}\left(\frac{1}{2},\frac{d+3}{2},\frac{d+5}{2},\frac{k_{R}^{2}}{4v^{2}}\right) & v>k_{R}/2
\end{cases},
\end{equation}
where $\,_{1}F_{2}\left(x\right)$ is Gauss's hypergeometric function.
For the special case of $d=2$,
\begin{equation}
D\left(v\right)=V_{0}^{2}\begin{cases}
6\pi\left(\frac{v}{k_{R}}\right)^{2} & v<k_{R}/2\\
\left(3\left(\frac{2v}{k_{R}}\right)^{2}\csc^{-1}\left(\frac{2v}{k_{R}}\right)-\left(\frac{k_{R}}{2v}\right)^{3}\left(2+3\left(\frac{2v}{k_{R}}\right)^{2}\right)\sqrt{1-\left(\frac{k_{R}}{2v}\right)^{2}}\right) & v>k_{R}/2
\end{cases}.\label{eq:Dv_2d_disk}
\end{equation}
The prediction of the FP equation was compared to the Monte-Carlo
simulation and is presented in Fig. \ref{fig:2d-cos-disk}. It is
found that uniform acceleration takes place for times $v<k_{R}/2$
and the FP prediction clearly fail in this regime. This can be expected
since from (\ref{eq:Dv_2d_disk}) $D\left(v\right)\sim v^{2}$, the
assumption that the velocity is constant (\ref{eq:FP_approx}) cannot
hold, rendering the FP approximation inconsistent. Nevertheless, for
$v>k_{R}/2$ the FP predictions are satisfied. Note that, for any
potential discussed in this section the scaling properties of Sec.
\ref{sec:Scaling-properties} hold.

Numerically, it is found that for large velocities ($v>k_{R}/2$)
the average square position is growing ballisticly, $\left\langle x^{2}\right\rangle \sim t^{2}$,
as expected from previous studies \cite{Rosenbluth1992,Golubovic1991,Bezuglyy2006,Bezuglyy2012},
while for small velocities ($v<k_{R}/2$), $\left\langle x^{2}\right\rangle \sim t^{4}$,
as expected for uniform acceleration.

\subsection{\label{sec:Potentials-Optics}Potentials proportional to the intensity
of a superposition of waves}

\begin{figure}
\begin{centering}
\includegraphics{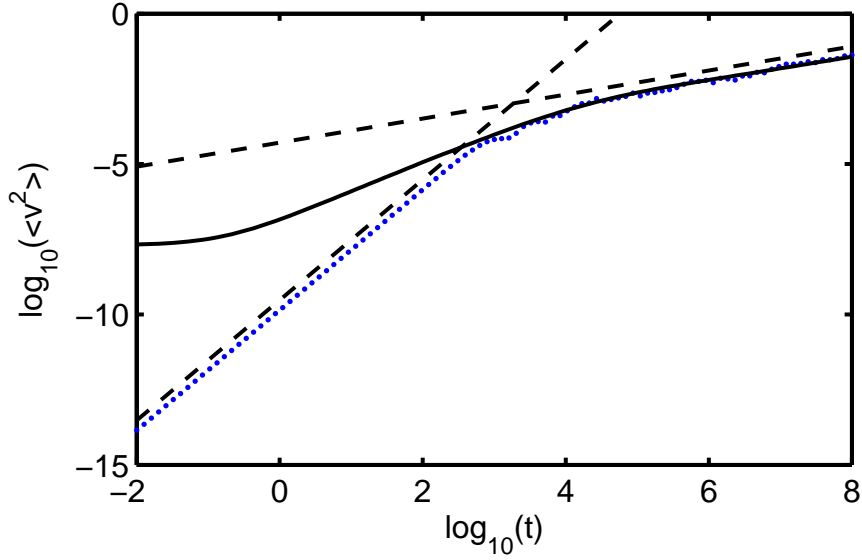}
\par\end{centering}

\caption{\label{fig:2d-Opt-disk}Same as Fig. \ref{fig:2d-cos-disk} but with
the potential (\ref{eq:Optic_potential}) and a uniform distribution
of wave-numbers given by (\ref{eq:fk_uniform_disk}). The dashed black
lines are guides for the eye with the corresponding slopes of $2$
and $2/5$. The parameters used for this simulation are, $I_{0}=10^{-4}$,
$k_{R}=0.1$.}
\end{figure}

In this subsection the diffusion coefficient for potentials of the
form (\ref{eq:2}) will be calculated. These potentials appear in
experiments with neutral atoms and in some experiments in optics,
where the refractive index (which plays the role of the potential)
of a photosensitive material is proportional to the intensity of light
\cite{Efremidis2002,Fleischer2003}. The complex field, $U$, which
represents the superposition pattern of waves is given by,

\begin{equation}
U\left(\mathbf{x},t\right)=\int\mathrm{d}\mathbf{k}\,\hat{U}\left(\mathbf{k}\right)\exp i\left(\mathbf{k}\cdot\mathbf{x}-\omega\left(k\right)t\right),
\end{equation}
where $\hat{U}\left(\mathbf{k}\right)$ are random Fourier coefficients,
and $\omega\left(k\right)$ is some dispersion relation. Following
the analysis of the previous subsection we will assume that $U$ is
composed of plane-waves with independent phases and amplitudes, leading
to
\begin{eqnarray}
\left\langle \hat{U}\left(\mathbf{k}\right)\right\rangle  & = & 0\label{eq:Optic_assump}\\
\left\langle \hat{U}\left(\mathbf{k}_{1}\right)\hat{U}^{*}\left(\mathbf{k}_{2}\right)\right\rangle  & =I_{0} & f\left(\mathbf{k}_{1}\right)\delta\left(\mathbf{k}_{1}-\mathbf{k}_{2}\right)\nonumber \\
\left\langle \hat{U}\left(\mathbf{k}_{1}\right)\hat{U}\left(\mathbf{k}_{2}\right)\right\rangle  & = & 0,\nonumber
\end{eqnarray}
where $\left\langle \left|\hat{U}\left(\mathbf{k}\right)\right|^{2}\right\rangle =I_{0}$
and $f\left(\mathbf{k}\right)$ is the probability density of the
wave numbers $\mathbf{k}$. The potential is proportional to the intensity
of the $U$,
\begin{equation}
V\left(\mathbf{x},t\right)=\left|U\left(\mathbf{x},t\right)\right|^{2}=\int\mathrm{d}\mathbf{k}_{1}\mathrm{d}\mathbf{k}_{2}\,\hat{U}\left(\mathbf{k}_{1}\right)\hat{U}^{*}\left(\mathbf{k}_{2}\right)\exp i\left[\left(\mathbf{k}_{1}-\mathbf{k}_{2}\right)\cdot\mathbf{x}-\left(\omega\left(k_{1}\right)-\omega\left(k_{2}\right)\right)t\right],\label{eq:Optic_potential}
\end{equation}
Using the assumptions (\ref{eq:Optic_assump}) we get that the potential
is constant on average,
\begin{eqnarray}
\left\langle V\left(\mathbf{x},t\right)\right\rangle  & = & \int\mathrm{d}\mathbf{k}_{1}\mathrm{d}\mathbf{k}_{2}\,\left\langle \hat{U}\left(\mathbf{k}_{1}\right)\hat{U}^{*}\left(\mathbf{k}_{2}\right)\right\rangle \exp i\left[\left(\mathbf{k}_{1}-\mathbf{k}_{2}\right)\cdot\mathbf{x}-\left(\omega\left(k_{1}\right)-\omega\left(k_{2}\right)\right)t\right]\nonumber \\
 & = & I_{0}\int\mathrm{d}\mathbf{k}\, f\left(\mathbf{k}\right)=I_{0}.
\end{eqnarray}
The correlation function of the potential is,
\begin{eqnarray}
\left\langle V\left(\mathbf{x}_{1},t_{1}\right)V\left(\mathbf{x}_{2},t_{2}\right)\right\rangle  & = & \int\mathrm{d}\mathbf{k}_{1}\mathrm{d}\mathbf{k}_{2}\mathrm{d}\mathbf{k}_{3}\mathrm{d}\mathbf{k}_{4}\,\left\langle \hat{U}\left(\mathbf{k}_{1}\right)\hat{U}^{*}\left(\mathbf{k}_{2}\right)\hat{U}\left(\mathbf{k}_{3}\right)\hat{U}^{*}\left(\mathbf{k}_{4}\right)\right\rangle \nonumber \\
 & \times & \exp i\left[\left(\mathbf{k}_{1}-\mathbf{k}_{2}\right)\cdot\mathbf{x}_{1}-\left(\omega\left(k_{1}\right)-\omega\left(k_{2}\right)\right)t_{1}\right]\nonumber \\
 & \times & \exp i\left[\left(\mathbf{k}_{3}-\mathbf{k}_{4}\right)\cdot\mathbf{x}_{2}-\left(\omega\left(k_{3}\right)-\omega\left(k_{4}\right)\right)t_{2}\right].
\end{eqnarray}
Since we have assumed that the complex amplitudes of the plane-waves
are independent random variables we can decompose,
\begin{eqnarray}
\left\langle \hat{U}\left(\mathbf{k}_{1}\right)\hat{U}^{*}\left(\mathbf{k}_{2}\right)\hat{U}\left(\mathbf{k}_{3}\right)\hat{U}^{*}\left(\mathbf{k}_{4}\right)\right\rangle  & = & \left\langle \hat{U}\left(\mathbf{k}_{1}\right)\hat{U}^{*}\left(\mathbf{k}_{2}\right)\right\rangle \left\langle \hat{U}\left(\mathbf{k}_{3}\right)\hat{U}^{*}\left(\mathbf{k}_{4}\right)\right\rangle \nonumber \\
 & + & \left\langle \hat{U}\left(\mathbf{k}_{1}\right)\hat{U}^{*}\left(\mathbf{k}_{4}\right)\right\rangle \left\langle \hat{U}^{*}\left(\mathbf{k}_{2}\right)\hat{U}\left(\mathbf{k}_{3}\right)\right\rangle \nonumber \\
 & = & I_{0}^{2}f\left(\mathbf{k}_{1}\right)f\left(\mathbf{k}_{3}\right)\delta\left(\mathbf{k}_{1}-\mathbf{k}_{2}\right)\delta\left(\mathbf{k}_{3}-\mathbf{k}_{4}\right)\nonumber \\
 & + & I_{0}^{2}f\left(\mathbf{k}_{1}\right)f\left(\mathbf{k}_{2}\right)\delta\left(\mathbf{k}_{1}-\mathbf{k}_{4}\right)\delta\left(\mathbf{k}_{2}-\mathbf{k}_{3}\right).
\end{eqnarray}
Therefore, the correlation function reduces to,
\begin{eqnarray}
C\left(\mathbf{x}_{1}-\mathbf{x}_{2},t_{1}-t_{2}\right) & = & I_{0}^{2}\\
 & + & I_{0}^{2}\int\mathrm{d}\mathbf{k}_{1}\mathrm{d}\mathbf{k}_{2}f\left(\mathbf{k}_{1}\right)f\left(\mathbf{k}_{2}\right)\nonumber \\
 & \times & \exp i\left[\left(\mathbf{k}_{1}-\mathbf{k}_{2}\right)\cdot\left(\mathbf{x}_{1}-\mathbf{x}_{2}\right)-\left(\omega\left(k_{1}\right)-\omega\left(k_{2}\right)\right)\left(t_{1}-t_{2}\right)\right].\nonumber
\end{eqnarray}
The PSD is just,
\begin{eqnarray}
S\left(\mathbf{q},\omega\right) & = & \frac{1}{\left(2\pi\right)^{d+1}}\int\mathrm{d}\mathbf{x}\int\mathrm{d}t\, C\left(\mathbf{x},t\right)\exp-i\left(\mathbf{q}\cdot\mathbf{x}-\omega t\right)\\
 & = & I_{0}^{2}\delta\left(\mathbf{q}\right)\delta\left(\omega\right)\nonumber \\
 & + & \frac{I_{0}^{2}}{\left(2\pi\right)^{d+1}}\int\mathrm{d}\mathbf{x}\int\mathrm{d}t\,\int\mathrm{d}\mathbf{k}_{1}\mathrm{d}\mathbf{k}_{2}f\left(\mathbf{k}_{1}\right)f\left(\mathbf{k}_{2}\right)\nonumber \\
 & \times & \exp i\left[\left(\mathbf{k}_{1}-\mathbf{k}_{2}-\mathbf{q}\right)\cdot\mathbf{x}-\left[\omega\left(k_{1}\right)-\omega\left(k_{2}\right)-\omega\right]t\right].
\end{eqnarray}
Taking the integral with respect to $\mathbf{x}$ and $t$ we get,
\begin{equation}
S\left(\mathbf{q},\omega\right)=I_{0}^{2}\delta\left(\mathbf{q}\right)\delta\left(\omega\right)+I_{0}^{2}\int\mathrm{d}\mathbf{k}_{1}\mathrm{d}\mathbf{k}_{2}f\left(\mathbf{k}_{1}\right)f\left(\mathbf{k}_{2}\right)\delta\left(\mathbf{k}_{1}-\mathbf{k}_{2}-\mathbf{q}\right)\delta\left[\omega-\left(\omega\left(k_{1}\right)-\omega\left(k_{2}\right)\right)\right],
\end{equation}
or
\begin{eqnarray}
S\left(\mathbf{q},\omega\right) & = & I_{0}^{2}\delta\left(\mathbf{q}\right)\delta\left(\omega\right)+I_{0}^{2}\int\mathrm{d}\mathbf{k}\, f\left(\mathbf{k}\right)f\left(\mathbf{k}-\mathbf{q}\right)\delta\left[\omega-\left(\omega\left(k\right)-\omega\left(\left|\mathbf{k}-\mathbf{q}\right|\right)\right)\right].
\end{eqnarray}
The diffusion coefficient is given by (see, (\ref{eq:Dv_d}))
\begin{equation}
D\left(v\right)=\pi I_{0}^{2}\int\mathrm{d}\mathbf{q}\mathrm{d}\mathbf{k}\,\left(\mathbf{q}\cdot\hat{v}\right)^{2}f\left(\mathbf{k}\right)f\left(\mathbf{k}-\mathbf{q}\right)\delta\left(\mathbf{q}\cdot\mathbf{v}-\left(\omega\left(k\right)-\omega\left(\left|\mathbf{k}-\mathbf{q}\right|\right)\right)\right).
\end{equation}
Changing the variables to,
\begin{eqnarray}
\mathbf{q}' & = & \mathbf{k}-\mathbf{q}\nonumber \\
\mathbf{k}' & = & \mathbf{k},
\end{eqnarray}
and suppressing the primes for simplicity gives,
\begin{equation}
D\left(v\right)=\pi I_{0}^{2}\int\mathrm{d}\mathbf{q}\mathrm{d}\mathbf{k}\,\left(\left(\mathbf{k}-\mathbf{q}\right)\cdot\hat{v}\right)^{2}f\left(\mathbf{k}\right)f\left(\mathbf{q}\right)\delta\left[\left(\mathbf{k}-\mathbf{q}\right)\cdot\mathbf{v}-\left(\omega\left(k\right)-\omega\left(q\right)\right)\right].\label{eq:84}
\end{equation}
Choosing the $x$- components of $\mathbf{k}$ and $\mathbf{q}$ such
that they are aligned with the velocity $v$ yields,
\begin{eqnarray}
D\left(v\right) & = & \pi I_{0}^{2}\left(\frac{S_{d}}{2\pi}\right)^{2}\int_{0}^{\infty}\mathrm{d}q\int_{0}^{2\pi}\mathrm{d}\theta_{q}\int_{0}^{\infty}\mathrm{d}k\int_{0}^{2\pi}\mathrm{d}\theta_{k}\,\left(qk\right)^{d-1}\label{eq:85}\\
 & \times & \left(k\cos\theta_{k}-q\cos\theta_{q}\right)^{2}f\left(k\right)f\left(q\right)\delta\left(v\left(k\cos\theta_{k}-q\cos\theta_{q}\right)-\left(\omega\left(k\right)-\omega\left(q\right)\right)\right).\nonumber
\end{eqnarray}
Changing variables to, $y_{k}=v\cos\theta_{k}$ and $y_{q}=v\cos\theta_{q}$
in (\ref{eq:85}) we have the Jacobian,
\begin{equation}
J=\frac{1}{v^{2}\sqrt{\left(1-\left(y_{k}/v\right)^{2}\right)\left(1-\left(y_{q}/v\right)^{2}\right)}}.
\end{equation}
In these variables the diffusion coefficient is,
\begin{eqnarray}
D\left(v\right) & = & 4\pi I_{0}^{2}\left(\frac{S_{d}}{2\pi}\right)^{2}\frac{1}{v^{4}}\int_{0}^{\infty}\mathrm{d}q\int_{0}^{\infty}\mathrm{d}k\int_{-v}^{v}\mathrm{d}y_{k}\int_{-v}^{v}\mathrm{d}y_{q}\,\left(qk\right)^{d-1}\\
 & \times & \frac{\left(ky_{k}-qy_{q}\right)^{2}}{\sqrt{\left(1-\left(y_{k}/v\right)^{2}\right)\left(1-\left(y_{q}/v\right)^{2}\right)}}f\left(k\right)f\left(q\right)\delta\left(\left(ky_{k}-qy_{q}\right)-\left(\omega\left(k\right)-\omega\left(q\right)\right)\right).\nonumber
\end{eqnarray}
For large velocities this gives,
\begin{equation}
D\left(v\right)\sim\frac{D_{3}}{v^{3}},
\end{equation}
with
\begin{equation}
D_{3}=\frac{4}{\pi}I_{0}^{2}S_{d}^{2}\int_{0}^{\infty}\mathrm{d}q\, q^{d-1}\int_{0}^{\infty}\mathrm{d}k\, k^{d-2}\left(\omega\left(k\right)-\omega\left(q\right)\right)^{2}f\left(k\right)f\left(q\right).
\end{equation}
Since $f\left(k\right)$ is non-negative the integral for $D_{3}$
does not vanish and for an appropriate choice of $f\left(k\right)$
it is convergent. The diffusion coefficient for zero velocity can
be calculated from (\ref{eq:85}) by setting $v=0$,
\begin{eqnarray}
D\left(0\right) & = & \pi I_{0}^{2}\left(\frac{S_{d}}{2\pi}\right)^{2}\int_{0}^{\infty}\mathrm{d}q\int_{0}^{2\pi}\mathrm{d}\theta_{q}\int_{0}^{\infty}\mathrm{d}k\int_{0}^{2\pi}\mathrm{d}\theta_{k}\,\left(qk\right)^{d-1}\\
 & \times & \left(k\cos\theta_{k}-q\cos\theta_{q}\right)^{2}f\left(k\right)f\left(q\right)\delta\left(\omega\left(k\right)-\omega\left(q\right)\right).\nonumber
\end{eqnarray}
Assuming that $\omega\left(k\right)$ is a monotonic function we have,
\begin{equation}
\delta\left(\omega\left(k\right)-\omega\left(q\right)\right)=\frac{1}{\left|\omega'\left(q\right)\right|}\delta\left(k-q\right),
\end{equation}
and therefore,
\begin{eqnarray}
D\left(0\right) & = & \pi I_{0}^{2}\left(\frac{S_{d}}{2\pi}\right)^{2}\int_{0}^{2\pi}\mathrm{d}\theta_{q}\int_{0}^{2\pi}\mathrm{d}\theta_{k}\,\left(\cos\theta_{k}-\cos\theta_{q}\right)^{2}\nonumber \\
 & \times & \int_{0}^{\infty}\mathrm{d}q\, q^{2d}\frac{f^{2}\left(q\right)}{\left|\omega'\left(q\right)\right|}.
\end{eqnarray}
After integrating over the angles,
\begin{equation}
D\left(0\right)=\pi I_{0}^{2}S_{d}^{2}\int_{0}^{\infty}\mathrm{d}q\, q^{2d}\frac{f^{2}\left(q\right)}{\left|\omega'\left(q\right)\right|},
\end{equation}
that is non-vanishing. Therefore, the potentials described in this
section are generic, namely, produce the universal behavior both for
small and large velocities. We will now demonstrate the calculation
for a particular choice of dispersion relation, $\omega\left(k\right)=k^{2}/2$
and a uniform distribution function $f\left(k\right)$ on a disk given
by (\ref{eq:fk_uniform_disk}) the integrals in (\ref{eq:85}) cannot
be taken explicitly, however we have calculated them numerically (see
Fig. \ref{fig:Dv_2d}) and the resulting dynamics is demonstrated
in Fig. \ref{fig:2d-Opt-disk}.
 Also here we found numerically that $\left\langle x^{2}\right\rangle \sim t^{2}$
for large $t$ and $\left\langle x^{2}\right\rangle \sim t^{4}$ for
small $t$.

For same dispersion relation and a Gaussian distribution of wave numbers,
\begin{equation}
f\left(\mathbf{k}\right)=\frac{1}{\left(2\pi k_{R}^{2}\right)^{d/2}}e^{-\frac{k^{2}}{2k_{R}^{2}}},\label{eq:fk_Gauss}
\end{equation}
using (\ref{eq:84}) we have,
\begin{equation}
D\left(v\right)=\frac{\pi I_{0}^{2}}{\left(2\pi k_{R}^{2}\right)^{d}}\frac{1}{2\pi}\int_{-\infty}^{\infty}\mathrm{d}\tau\int\mathrm{d}\mathbf{q}\mathrm{d}\mathbf{k}\,\left(\left(\mathbf{k}-\mathbf{q}\right)\cdot\hat{v}\right)^{2}e^{-\frac{k^{2}+q^{2}}{2k_{R}^{2}}}\exp i\tau\left[\left(\mathbf{k}-\mathbf{q}\right)\cdot\mathbf{v}-\frac{1}{2}\left(k^{2}-q^{2}\right)\right],
\end{equation}
where we have used the exponential representation of the delta function.
Noticing that,
\begin{equation}
D\left(v\right)=\frac{\pi I_{0}^{2}}{\left(2\pi k_{R}^{2}\right)^{d}}\frac{1}{2\pi}\left[\int_{-\infty}^{\infty}-\frac{\mathrm{d}\tau}{\tau^{2}}\int\mathrm{d}\mathbf{q}\mathrm{d}\mathbf{k}\, e^{-\frac{k^{2}+q^{2}}{2k_{R}^{2}}}\frac{\partial^{2}}{\partial v^{2}}\exp i\tau\left[\left(\mathbf{k}-\mathbf{q}\right)\cdot\mathbf{v}-\frac{1}{2}\left(k^{2}-q^{2}\right)\right]\right]
\end{equation}
and taking the Gaussian integrals gives,
\begin{equation}
D\left(v\right)=\frac{\pi I_{0}^{2}}{\left(2\pi k_{R}^{2}\right)^{d}}\frac{1}{2\pi}\left[\int_{-\infty}^{\infty}-\frac{\mathrm{d}\tau}{\tau^{2}}\frac{\left(2\pi k_{R}^{2}\right)^{d}}{\left(1+k_{R}^{4}\tau^{2}\right)^{d/2}}\frac{\partial^{2}}{\partial v^{2}}e^{-\left(\frac{k_{R}^{2}\tau^{2}}{1+k_{R}^{4}\tau^{2}}\right)v^{2}}\right],
\end{equation}
or
\begin{equation}
D\left(v\right)=\pi I_{0}^{2}\left[\frac{1}{2\pi}\int_{-\infty}^{\infty}-\frac{\mathrm{d}\tau}{\tau^{2}}\frac{1}{\left(1+k_{R}^{4}\tau^{2}\right)^{d/2}}\frac{\partial^{2}}{\partial v^{2}}e^{-\left(\frac{k_{R}^{2}\tau^{2}}{1+k_{R}^{4}\tau^{2}}\right)v^{2}}\right].\label{eq:Dv_2d_opt_Gauss}
\end{equation}
The dynamics following from (\ref{eq:Dv_2d_opt_Gauss}) is rather
similar to that demonstrated in Fig. \ref{fig:2d-Opt-disk} and therefore
it is not presented here.

\section{\label{sec:Anomalous-diffusion-1D}Anomalous diffusion in one dimensional
systems}

In this section we will consider one dimensional systems and explore
the different possibilities of asymptotic behavior of the velocity
distribution. In particular, we will show that the asymptotic behavior
of the form (\ref{eq:D_univ}), is \emph{not} universal since some
possibilities were not considered in previous studies \cite{Golubovic1991,Bezuglyy2006,Bezuglyy2012}.
In the first subsection we will present a some representative examples
of the new universality classes, and in the second subsection we will
explain when the asymptotic expansion of \cite{Golubovic1991,Bezuglyy2006,Bezuglyy2012}
is not effective, and provide a prescription for designing transport
properties using the PSD.

\subsection{New universality classes}

Following from (\ref{eq:Dv_d}) the diffusion coefficient for a one
dimensional system takes the form,
\begin{equation}
D\left(v\right)=\pi\int k^{2}S\left(k,kv\right)\mathrm{d}k.\label{eq:Dv_1d}
\end{equation}
For a PSD with the property,
\begin{equation}
S\left(k,\omega\right)=0\qquad\omega/k>v_{\max},\label{eq:S_bounded}
\end{equation}
implies that the diffusion coefficient is zero for large velocities,
\begin{equation}
D\left(v\right)=0\qquad v>v_{\max},\label{eq:D_1d_zero}
\end{equation}
as demonstrated in \cite{Krivolapov2012,Krivolapov2012a}. The correlation
function corresponding to the PSD with the property (\ref{eq:S_bounded})
may be infinitely differentiable, nevertheless the diffusion coefficient
vanishes for large velocities as is clear from (\ref{eq:D_1d_zero}),
a possibility which was overlooked in \cite{Golubovic1991,Bezuglyy2006,Bezuglyy2012}.

We will now present specific examples of this new universality class
using the specific potentials of the form (\ref{eq:1}) and (\ref{eq:2}).
The PSD of potentials of the type (\ref{eq:1}) is,
\begin{equation}
S\left(k,\omega\right)=V_{0}^{2}f\left(k\right)\left[\delta\left(\omega-\omega\left(k\right)\right)+\delta\left(\omega+\omega\left(k\right)\right)\right].\label{eq:Sk_cos}
\end{equation}
Taking the dispersion relation $\omega=k^{2}/2$, and using (\ref{eq:Dv_1d})
and (\ref{eq:Sk_cos}) yields,
\begin{eqnarray}
D\left(v\right) & = & \pi V_{0}^{2}\int k^{2}f\left(k\right)\left[\delta\left(kv-k^{2}/2\right)+\delta\left(kv+k^{2}/2\right)\right]\mathrm{d}k\nonumber \\
 & = & \pi V_{0}^{2}\int k^{2}f\left(k\right)\left[\frac{\delta\left(k\right)}{\left|v\right|}+\frac{\delta\left(k-2v\right)}{\left|v\right|}+\frac{\delta\left(k+2v\right)}{\left|v\right|}\right]\mathrm{d}k,
\end{eqnarray}
which after integrating over $k$ gives,
\begin{equation}
D\left(v\right)=8\pi V_{0}^{2}\left|v\right|f\left(2v\right).\label{eq:Dv_cos_quad}
\end{equation}
Any asymptotic behavior is possible as the diffusion coefficient is
proportional to $f\left(v\right)$ and therefore clearly may decay
faster than $v^{-3}$. The correlation function for these potentials
can be obtained for example for a Gaussian distribution of wave-numbers
(\ref{eq:fk_Gauss}). Using (\ref{eq:Weiner-Khintchin}) and (\ref{eq:Sk_cos})
we have,
\begin{equation}
C\left(x,t\right)=2V_{0}^{2}\frac{\cos\left[\frac{k_{R}^{4}t\, x^{2}}{2\left(1+k_{R}^{4}t^{2}\right)}-\frac{1}{2}\arg\left(1+ik_{R}^{2}t\right)\right]}{\left(1+k_{R}^{4}t^{2}\right)^{1/4}}\exp\left[-\frac{k_{R}^{2}x^{2}}{2\left(1+k_{R}^{4}t^{2}\right)}\right].\label{eq:Cxt_cos_Gauss}
\end{equation}
Note, that this correlation function is an infinitely differentiable
function over the whole $\left(x,t\right)$ plane. It has fast decaying
spatial correlations compared to a very slow temporal correlations.

\begin{figure}
\begin{centering}
\includegraphics{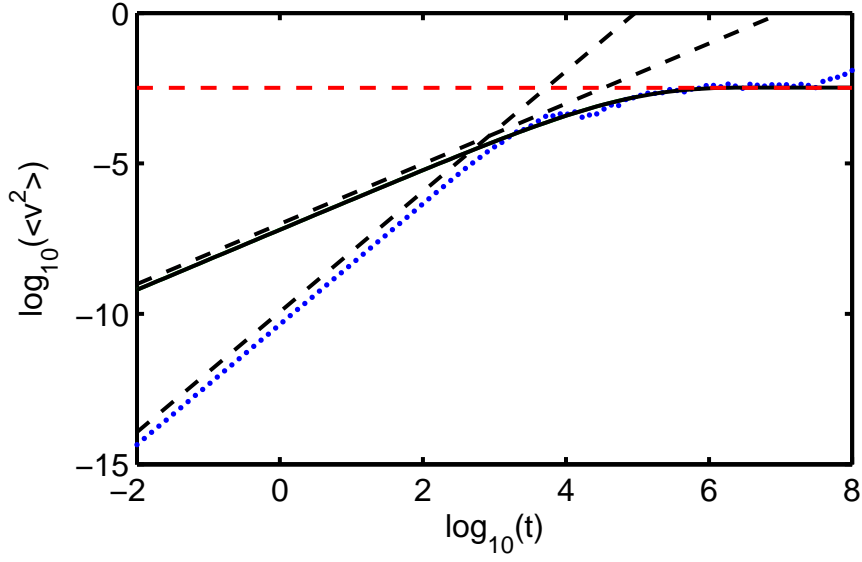}
\par\end{centering}

\caption{\label{fig:1d-opt-disk}A log-log plot of average squared velocity
as a function of time for a one dimensional system with a potential
(\ref{eq:Optic_potential}) and a distribution of wave-numbers (\ref{eq:fk_uniform_disk}).
The blue dots represent the result of the Monte-Carlo solution of
(\ref{eq:Hamilton_eq}) averaged over $20$ realizations and the black
solid line is the numerical solution of the Fokker-Planck equation
for the velocity (\ref{eq:FP_momentum}). The dashed black and red
lines are guides for the eye with the corresponding slopes of 2 ,
1 and 0. The initial condition was a narrow distribution of velocities
for the Fokker-Planck and $x=v=0$ for the Monte-Carlo calculation.
The parameters used for this simulation are, $I_{0}=10^{-4}$, $k_{R}=0.1$.}
\end{figure}

For potentials of the type (\ref{eq:2}) the PSD is given by,
\begin{equation}
S\left(k,\omega\right)=I_{0}^{2}\int\mathrm{d}q\, f\left(q\right)f\left(q-k\right)\delta\left[\omega-\left(\omega\left(q\right)-\omega\left(q-k\right)\right)\right].\label{eq:Sk_optic}
\end{equation}
As in the preceding example, taking a quadratic dispersion relation,
$\omega=k^{2}/2$, gives
\begin{eqnarray}
D\left(v\right) & = & \pi I_{0}^{2}\int k^{2}f\left(q\right)f\left(q-k\right)\delta\left[kv-\frac{1}{2}\left(q^{2}-\left(q-k\right)^{2}\right)\right]\mathrm{d}q\mathrm{d}k\nonumber \\
 & = & \pi I_{0}^{2}\int k^{2}f\left(q\right)f\left(q-k\right)\left[\frac{\delta\left(k\right)}{\left|v-q\right|}+\frac{\delta\left(k-2\left(q-v\right)\right)}{\left|v-q\right|}\right]\mathrm{d}q\mathrm{d}k\nonumber \\
 & = & 4\pi I_{0}^{2}\int\left|q-v\right|f\left(q\right)f\left(2v-q\right)\mathrm{d}q.
\end{eqnarray}
Changing the variables to $q=q'+v$ and suppressing the prime we have,
\begin{equation}
D\left(v\right)=4\pi I_{0}^{2}\int\left|q\right|f\left(q+v\right)f\left(q-v\right)\mathrm{d}q.\label{eq:Dv_opt_quad}
\end{equation}
And the decay of the diffusion coefficient with the velocity is also
dictated by the decay of $f\left(q\right)$. An explicit expression
may be obtained for example for a uniform distribution of wave-numbers
(\ref{eq:fk_uniform_disk}), giving \cite{Krivolapov2012,Krivolapov2012a},
\begin{equation}
D\left(v\right)=\begin{cases}
\frac{\pi I_{0}^{2}}{k_{R}^{2}}\left(k_{R}-\left|v\right|\right)^{2} & \left|v\right|<k_{R}\\
0 & \left|v\right|>k_{R}
\end{cases}.\label{eq:Dv_1d_segment}
\end{equation}
The resulting dynamics is demonstrated in Fig. \ref{fig:1d-opt-disk}.
The regime of a unit slope corresponds to regular diffusion in velocity
and the asymptotic regime corresponds to an absence of diffusion in
velocity. We have verified numerically that the asymptotic regime
corresponds to a ballistic growth in position, $\left\langle x^{2}\right\rangle \sim t^{2}$.

\begin{figure}
\begin{centering}
\includegraphics{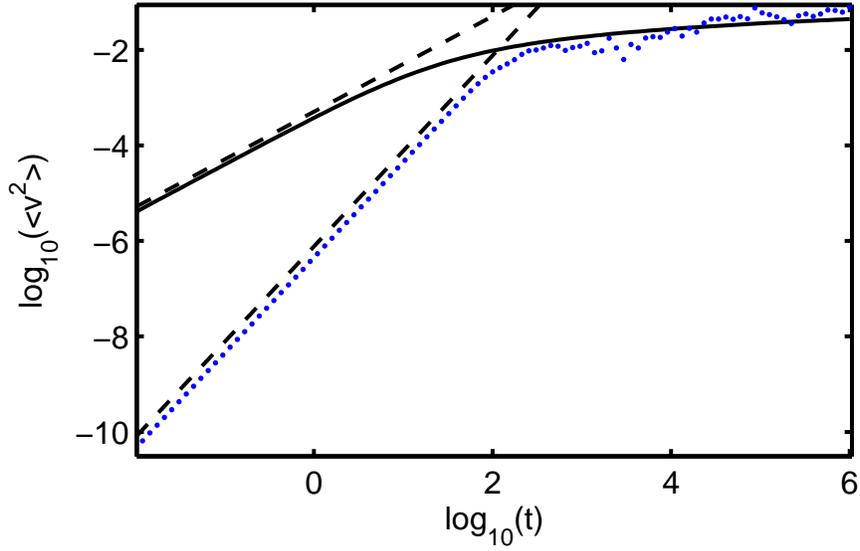}
\par\end{centering}

\caption{\label{fig:1d-opt-Gauss}Same as Fig. \ref{fig:1d-opt-disk} but with
a distribution of wave-numbers (\ref{eq:fk_Gauss}). The dashed black
lines are guides for the eye with the corresponding slopes of 2 and
1. The parameters used for this simulation are, $V_{0}=10^{-2}$,
$k_{R}=0.1$.}
\end{figure}

Another explicit expression may be obtained for a Gaussian distribution
of wave-numbers (\ref{eq:fk_Gauss}). The diffusion coefficient is
given by,
\begin{equation}
D\left(v\right)=2I_{0}^{2}e^{-\frac{v^{2}}{k_{R}^{2}}}.\label{eq:Dv_1d_opt_Gauss}
\end{equation}
The predictions of this equation are compared with Monte-Carlo simulation
and the results are presented in Fig. \ref{fig:1d-opt-Gauss}. The
correlation function for this potentials can be obtained using (\ref{eq:Weiner-Khintchin}),
\begin{equation}
C\left(x,t\right)=I_{0}^{2}\frac{1}{\sqrt{1+k_{R}^{4}t^{2}}}\exp\left(-\frac{k_{R}^{2}x^{2}}{1+k_{R}^{4}t^{2}}\right).\label{eq:Cxt_opt_Gauss}
\end{equation}
Similarly, to the correlation function (\ref{eq:Cxt_cos_Gauss}) this
correlation function is infinitely differentiable, decaying fast in
position and slowly decaying in time.

This existence of a dispersion relation between the spatial and temporal
frequencies is not a necessary condition for the new universality
class, as is demonstrated by the following PSD,
\begin{equation}
S\left(k,\omega\right)=\frac{eV_{0}^{2}}{4\pi k_{0}\omega_{0}}e^{-\left(\frac{k_{0}}{k}\right)^{2}}e^{-\left(\frac{\omega^{2}}{2\omega_{0}^{2}}+\frac{k^{2}}{2k_{0}^{2}}\right)},\label{eq:Sk_essen_sing}
\end{equation}
where the normalization factor was chosen such that
\begin{equation}
\int S\left(k,\omega\right)\mathrm{d}k\mathrm{d}\omega=V_{0}^{2}.
\end{equation}
Using (\ref{eq:Dv_1d}) and defining $v_{0}=\omega_{0}/k_{0}$ yields,
\begin{equation}
D\left(v\right)=\frac{eV_{0}^{2}}{\sqrt{\pi}}\frac{k_{0}}{v_{0}}\left[\frac{1}{\left(1+\left(v/v_{0}\right)^{2}\right)^{3/2}}+\frac{1}{\left(1+\left(v/v_{0}\right)^{2}\right)}\right]e^{-\sqrt{1+\left(v/v_{0}\right)^{2}}},\label{eq:Dv_essent_sing}
\end{equation}
which is exponentially decaying for a large velocity. We could not
obtain an analytical expression for the correlation function, but
it can be expressed as an integral,
\begin{eqnarray}
C\left(x,t\right) & = & \frac{eV_{0}^{2}}{4\pi k_{0}\omega_{0}}\int e^{-\left(\frac{k_{0}}{k}\right)^{2}}e^{-\left(\frac{\omega^{2}}{2\omega_{0}^{2}}+\frac{k^{2}}{2k_{0}^{2}}\right)}\exp i\left(kx-\omega t\right)\mathrm{d}k\mathrm{d}\omega,\nonumber \\
 & = & \frac{eV_{0}^{2}}{2k_{0}\sqrt{2\pi}}e^{-\frac{t^{2}\omega_{0}^{2}}{2}}\int\exp\left(-\frac{k^{2}}{2k_{0}^{2}}-\frac{k_{0}^{2}}{k^{2}}\right)e^{ikx}\mathrm{d}k.\label{eq:Cxt_essnt_sing}
\end{eqnarray}
The temporal correlations are rapidly decaying and since the function
in the integrand is infinitely differentiable and integrable following
from the Riemann\textendash{}Lebesgue lemma the spatial correlations
will also decay faster than any power law.

In summary, in this subsection we have presented a few examples of
potentials, which give diffusion coefficients with various asymptotic
behavior for large velocities different from (\ref{eq:D_univ}). We
have also calculated their corresponding correlation functions and
showed that they are smooth and differentiable. Therefore, this class
of potentials spans a new universality class, which is qualitatively
different from the universal behavior described in previous studies
\cite{Golubovic1991,Bezuglyy2006,Bezuglyy2012}. We have also demonstrated
that the conditions for appearance of this class are not necessarily
connected to long range correlations or the special form of the potentials
(\ref{eq:1}) and (\ref{eq:2}). In particular, the existence of a
dispersion relation is \emph{not} required. In the next subsection
we will show that the conclusions drawn here are valid for any physical
dispersion relation, and present a classification of the potentials,
based on their PSDs, into different universality classes.

\subsection{Classification of potentials into different universality classes}

In the present work we have chosen to work with the PSD and not the
correlation function, since in relevant experiments \cite{Schwartz2007,Levi2011,Levi2011a}
the PSD rather then the correlation function is naturally controlled.
For the purpose of clarity we will repeat here the asymptotic expansion
of the diffusion coefficient, which was done in \cite{Bezuglyy2006,Bezuglyy2012}
and connect it to our representation using the PSD.

The PSD is the Fourier transform of the correlation function of the
potential, namely,
\begin{equation}
S\left(k,\omega\right)=\frac{1}{\left(2\pi\right)^{2}}\int C\left(x,t\right)\exp-i\left(kx-\omega t\right)\mathrm{d}x\mathrm{d}t.\label{eq:Win-Khin}
\end{equation}
Therefore using (\ref{eq:Dv_1d}),
\begin{equation}
D\left(v\right)=\frac{1}{4\pi}\int k^{2}C\left(x,t\right)\exp-ik\left(x-vt\right)\mathrm{d}x\mathrm{d}t\mathrm{d}k,
\end{equation}
which could be written as,
\begin{equation}
D\left(v\right)=-\frac{1}{4\pi}\int C\left(x,t\right)\frac{\partial^{2}}{\partial x^{2}}\exp-ik\left(x-vt\right)\mathrm{d}x\mathrm{d}t\mathrm{d}k.
\end{equation}
Integrating by parts twice and taking into the account that $C\left(x,t\right)$
and its derivatives vanish in the limit of $x\to\pm\infty$, we get,
\begin{equation}
D\left(v\right)=-\frac{1}{4\pi}\int\frac{\partial^{2}C}{\partial x^{2}}\left(x,t\right)\exp-ik\left(x-vt\right)\mathrm{d}x\mathrm{d}t\mathrm{d}k.
\end{equation}
Finally integrating over $k$ gives,
\begin{equation}
D\left(v\right)=-\frac{1}{2}\int\frac{\partial^{2}C}{\partial x^{2}}\left(x,t\right)\delta\left(x-vt\right)\mathrm{d}x\mathrm{d}t.
\end{equation}
This is the same expression for the diffusion coefficient, which was
obtained in previous studies (see Eq. 7 of \cite{Bezuglyy2006}).
For simplicity of notation we will set,
\begin{equation}
W\left(x,\tau\right)=\frac{\partial^{2}C}{\partial x^{2}}\left(x,\tau\right).\label{eq:Def_K}
\end{equation}
Changing variables to $t=t'/v$ we get,
\begin{eqnarray}
D\left(v\right) & = & -\frac{1}{2v}\int W\left(x,\frac{t'}{v}\right)\delta\left(x-t'\right)\mathrm{d}x\mathrm{d}t'\nonumber \\
 & = & -\frac{1}{2v}\int W\left(t',\frac{t'}{v}\right)\mathrm{d}t'.
\end{eqnarray}
Therefore, if $W$ is sufficiently differentiable, the asymptotic
expansion of $D\left(v\right)$ is,
\begin{equation}
D\left(v\right)=\frac{D_{1}}{v}+\frac{D_{2}}{v^{2}}+\frac{D_{3}}{v^{3}}\cdots,\label{eq:asympt_expan}
\end{equation}
where
\begin{equation}
D_{n}=-\frac{1}{2n!}\int x{}^{n-1}\frac{\partial^{n-1}W}{\partial\tau^{n-1}}\left(x,\tau\right)|_{\tau=0}\mathrm{d}x\qquad n=1,2,\ldots,.\label{eq:Dn_asympt}
\end{equation}
The first term vanishes, since
\begin{equation}
D_{1}=-\frac{1}{2}\int W\left(x,0\right)\mathrm{d}x=-\frac{1}{2}\int\frac{\partial^{2}C}{\partial x^{2}}\left(x,0\right)\mathrm{d}x=0,
\end{equation}
and all the even terms vanish due to the symmetry, $x\to-x$ (following
from the fact that the potential is translationaly invariant),
\begin{equation}
D_{2n}=-\frac{1}{2}\frac{1}{\left(2n\right)!}\int x{}^{2n-1}\frac{\partial^{2n-1}W}{\partial\tau^{2n-1}}\left(x,\tau\right)|_{\tau=0}\mathrm{d}x=0.
\end{equation}
Therefore, the leading term in the asymptotic expansion of the diffusion
coefficient is,
\begin{equation}
D\left(v\right)\sim\frac{D_{3}}{v^{3}},\label{eq:Dv_univ1}
\end{equation}
where
\begin{equation}
D_{3}=-\frac{1}{12}\int x{}^{2}\frac{\partial^{2}W}{\partial\tau^{2}}\left(x,\tau\right)|_{\tau=0}\mathrm{d}x.
\end{equation}
In \cite{Golubovic1991,Bezuglyy2006,Bezuglyy2012} it was claimed
that if the correlation function is at least twice differentiable
in both arguments then the asymptotic behavior (\ref{eq:Dv_univ1})
follows, and furthermore it is universal. However, as was demonstrated
in the previous subsection exceptions are possible, since some or
even \emph{all} of the terms $D_{n>3}$ may vanish, giving a qualitatively
different asymptotic behavior. If the diffusion coefficient vanishes
or decreases faster than any power law for large velocities the asymptotic
expansion in powers of $v^{-1}$ clearly vanishes. Thus the transport
in such case is sub-diffusive for large velocity and the growth of
the moments of the velocity distribution is slower than any power
law. In what follows we will demonstrate that for the potentials of
type (\ref{eq:1}) and (\ref{eq:2}) the asymptotic expansion of $D\left(v\right)$
vanishes, for any choice of the dispersion relation, provided that
the distribution of the wave-numbers decays faster than any power
law.

Taking (\ref{eq:Sk_cos}) and the relation (\ref{eq:Win-Khin}) we
can calculate,
\begin{eqnarray}
W\left(x,t\right) & = & -V_{0}^{2}\int k^{2}f\left(k\right)\left[\delta\left(\omega-\omega\left(k\right)\right)+\delta\left(\omega+\omega\left(k\right)\right)\right]\exp i\left(kx-\omega t\right)\mathrm{d}k\mathrm{d}\omega\nonumber \\
 & = & -V_{0}^{2}\int k^{2}f\left(k\right)\left[\exp i\left(kx-\omega\left(k\right)t\right)+\exp i\left(kx+\omega\left(k\right)t\right)\right]\mathrm{d}k.
\end{eqnarray}
Using the expression for the coefficients of the asymptotic expansion
(\ref{eq:Dn_asympt}) we have,
\begin{eqnarray}
D_{n} & = & \frac{V_{0}^{2}}{n!}\int k^{2}f\left(k\right)\left(-i\omega\left(k\right)\right)^{n-1}t'^{n-1}e^{ikt'}\mathrm{d}k\mathrm{d}t'\nonumber \\
 & = & \frac{V_{0}^{2}}{n!}\int k^{2}f\left(k\right)\omega^{n-1}\left(k\right)\frac{\partial^{n-1}}{\partial k^{n-1}}e^{ikt'}\mathrm{d}k\mathrm{d}t'.
\end{eqnarray}
Integrating by parts $n-1$ times, and assuming that $f\left(k\right)$
decreases faster than any power law, so that the boundary terms may
be disregarded, we obtain,
\begin{eqnarray*}
D_{n} & = & \frac{\left(-1\right)^{2n-1}V_{0}^{2}}{n!}\int e^{ikt'}\frac{\partial^{n-1}}{\partial k^{n-1}}\left(k^{2}f\left(k\right)\omega^{n-1}\left(k\right)\right)\mathrm{d}k\mathrm{d}t'\\
 & = & 2\pi\frac{\left(-1\right)^{2n-1}V_{0}^{2}}{n!}\left[\frac{\partial^{n-1}}{\partial k^{n-1}}\left(k^{2}f\left(k\right)\omega^{n-1}\left(k\right)\right)\right]_{k=0}.
\end{eqnarray*}
For a physical dispersion relation which has the property $\omega\left(0\right)=0$
we will have $D_{n}=0$ for all $n$. As was shown in (\ref{eq:Dv_cos_quad})
the diffusion coefficient in this case is not zero, but a function
which decays with velocity faster than any power law, rendering the
expansion (\ref{eq:asympt_expan}) useless. For a potential with a
PSD given by (\ref{eq:Sk_optic}) we have,
\begin{eqnarray}
W\left(x,t\right) & = & -I_{0}^{2}\int k^{2}f\left(q\right)f\left(k-q\right)\delta\left[\omega-\left(\omega\left(q\right)-\omega\left(q-k\right)\right)\right]\exp i\left(kx-\omega t\right)\mathrm{d}q\mathrm{d}k\mathrm{d}\omega\nonumber \\
 & = & -I_{0}^{2}\int k^{2}f\left(q\right)f\left(k-q\right)\exp i\left(kx-\left(\omega\left(q\right)-\omega\left(q-k\right)\right)t\right)\mathrm{d}q\mathrm{d}k.
\end{eqnarray}
The coefficients of the asymptotic expansion are therefore given by,
\begin{eqnarray}
D_{n} & = & \frac{I_{0}^{2}}{2n!}\int\left(-i\left(\omega\left(q\right)-\omega\left(q-k\right)\right)\right)^{n-1}k^{2}f\left(q\right)f\left(k-q\right)t'^{n-1}e^{ikt'}\mathrm{d}t'\mathrm{d}q\mathrm{d}k\nonumber \\
 &  & \frac{I_{0}^{2}}{2n!}\int\left(-i\left(\omega\left(q\right)-\omega\left(q-k\right)\right)\right)^{n-1}k^{2}f\left(q\right)f\left(k-q\right)\left(-i\frac{\partial}{\partial k}\right)^{n-1}e^{ikt'}\mathrm{d}t'\mathrm{d}q\mathrm{d}k.
\end{eqnarray}
Integrating by parts and assuming $f\left(k\right)$ decays faster
than any power law gives,
\begin{eqnarray}
D_{n} & = & I_{0}^{2}\frac{\left(-1\right)^{2n-1}}{2n!}\int e^{ikt'}\frac{\partial^{n-1}}{\partial k^{n-1}}\left\{ \left[\omega\left(q\right)-\omega\left(q-k\right)\right]^{n-1}k^{2}f\left(q\right)f\left(k-q\right)\right\} \mathrm{d}t'\mathrm{d}q\mathrm{d}k\nonumber \\
 & = & 2\pi I_{0}^{2}\frac{\left(-1\right)^{2n-1}}{2n!}\frac{\partial^{n-1}}{\partial k^{n-1}}\left\{ \int\left[\omega\left(q\right)-\omega\left(q-k\right)\right]^{n-1}k^{2}f\left(q\right)f\left(k-q\right)\mathrm{d}q\right\} _{k=0}.
\end{eqnarray}
These coefficients vanish for any dispersion relation, and the diffusion
coefficient is different from zero, and was obtained explicitly for
a specific choice of the dispersion relation in (\ref{eq:Dv_opt_quad}).

In experiments one usually does not control the diffusion coefficient
directly, and therefore we will classify the potentials leading to
slow transport using the PSD, which can be controlled. Starting from
a relation for the diffusion coefficient (\ref{eq:Dv_1d}) we change
the variable to $k=k'/v$ and obtain,
\begin{equation}
D\left(v\right)=\frac{\pi}{v^{3}}\int k'^{2}S\left(\frac{k'}{v},k'\right)\mathrm{d}k'.
\end{equation}
If the PSD is not singular (note, that this is not the case for the
potentials of type (\ref{eq:1}) and (\ref{eq:2}) discussed earlier)
and differentiable on the line $k=0$ (where $k$ is the first argument
of $S\left(k,\omega\right))$ we can expand,
\begin{equation}
D\left(v\right)=\frac{D_{3}}{v^{3}}+\frac{D_{5}}{v^{5}}+\cdots,\label{eq:Sk_Asympt_exp}
\end{equation}
where
\begin{equation}
D_{n}=\frac{1}{\left(n-3\right)!}\int\mathrm{d}\omega\,\omega{}^{n-1}\frac{\partial^{n-3}S}{\partial k^{n-3}}\left(k,\omega\right)|_{k=0}\qquad n\geq3.
\end{equation}
Note, that in this representation the leading order of the expansion
is directly $v^{-3},$ since representation using the PSD already
contains the assumption that the potential is stationary. If the first
$n_{\max}$ derivatives, $\partial^{n}S/\partial k^{n}$, vanish on
the line $k=0$, where $n_{\max}$ is the first derivative that is
different from zero, then the resulting asymptotic behavior is $D\left(v\right)\sim D_{n_{\max}+3}/v^{n_{\max}+3}$.
\emph{All} the derivatives vanish only if $S\left(k,\omega\right)$
is non-analytic on the line $k=0$. The non-analytic behavior may
result either from a function, which is strictly zero on some finite
strip around the line $k=0$, as in (\ref{eq:S_bounded}) or due to
an essential singularity of $S\left(k,\omega\right)$ on this line,
which will be demonstrated bellow. The first case will lead to a diffusion
coefficient that will vanish for large velocities, for example (\ref{eq:Dv_cos_quad})
(with finite support $f\left(k\right)$) or (\ref{eq:Dv_1d_segment}).
The second case will result in a decay faster than any power law,
for example (\ref{eq:Dv_1d_opt_Gauss}). In experiments with a good
control of the PSD, by controlling the amount of vanishing derivatives
of the PSD on the line $k=0$, one can vary the asymptotic behavior
of the velocity distribution. The range of variation is from $D\left(v\right)\sim v^{-3}$
, through $D\left(v\right)\sim v^{-n}$ (with $n>3$), sub-exponential,
exponential, super-exponential and up-to $D\left(v\right)=0$ (for
$v>v_{\max}$).

We will now turn to examine the correlation function of the potential,
$C\left(x,t\right)$, for potentials, which belong to the new universality
classes. For simplicity we will consider the behavior along the line,
\begin{equation}
c\left(t\right)\equiv C\left(0,t\right)=\int\left(\int S\left(k,\omega\right)\mathrm{d}k\right)e^{-i\omega t}\mathrm{d}\omega
\end{equation}
 and,
\begin{equation}
c\left(x\right)\equiv C\left(x,0\right)=\int\left(\int S\left(k,\omega\right)\mathrm{d}\omega\right)e^{ikx}\mathrm{d}k.
\end{equation}
The moments of the correlation function along these lines are,
\begin{equation}
\left\langle t^{n}\right\rangle =\int t^{n}c\left(t\right)\mathrm{d}t
\end{equation}
and
\begin{equation}
\left\langle x^{n}\right\rangle =\int x^{n}c\left(x\right)\mathrm{d}x.
\end{equation}
For the PSD (\ref{eq:Sk_cos}) we have,
\begin{eqnarray}
\left\langle t^{n}\right\rangle  & = & V_{0}^{2}\int t^{n}f\left(k\right)\left[\delta\left(\omega-\omega\left(k\right)\right)+\delta\left(\omega+\omega\left(k\right)\right)\right]e^{-i\omega t}\mathrm{d}k\mathrm{d}\omega\mathrm{d}t\nonumber \\
 & = & V_{0}^{2}\int f\left(k\right)t^{n}\left(e^{-i\omega\left(k\right)t}+e^{i\omega\left(k\right)t}\right)\mathrm{d}k\mathrm{d}t.
\end{eqnarray}
We will now calculate the second moment, which will contain all the
representative features. Using the fact that,
\begin{equation}
\left\langle t^{2}\right\rangle =2V_{0}^{2}\int f\left(k\right)\left(-\frac{\partial^{2}}{\partial\omega^{2}}\right)\cos\omega t\,\mathrm{d}k\mathrm{d}t,
\end{equation}
and
\begin{equation}
\frac{\partial}{\partial\omega}=\frac{\mathrm{d}k}{\mathrm{d}\omega}\frac{\partial}{\partial k},
\end{equation}
we have,
\begin{equation}
\left\langle t^{2}\right\rangle =2V_{0}^{2}\int f\left(k\right)\left(\frac{\mathrm{d}k}{\mathrm{d}\omega}\frac{\partial}{\partial k}\frac{\mathrm{d}k}{\mathrm{d}\omega}\frac{\partial}{\partial k}\right)\cos\omega t\,\mathrm{d}k\mathrm{d}t.
\end{equation}
Integrating by parts twice gives,
\begin{eqnarray}
\left\langle t^{2}\right\rangle  & = & 2V_{0}^{2}\int\left(\frac{\partial}{\partial k}\frac{\mathrm{d}k}{\mathrm{d}\omega}\frac{\partial}{\partial k}\left(\frac{\mathrm{d}k}{\mathrm{d}\omega}f\left(k\right)\right)\right)\cos\omega t\,\mathrm{d}k\mathrm{d}t.\nonumber \\
 & = & 4\pi V_{0}^{2}\left[\left(\frac{\partial}{\partial k}\frac{\mathrm{d}k}{\mathrm{d}\omega}\frac{\partial}{\partial k}\left(\frac{\mathrm{d}k}{\mathrm{d}\omega}f\left(k\right)\right)\right)\right]_{k=0},
\end{eqnarray}
where we have used the assumption that $\omega\left(k\right)=0$ if
and only if $k=0$. Note, that for any increasing (near $k=0$) dispersion
relation, except $\omega=\left|k\right|c$, this moment diverges.
For the later dispersion relation the derivative, $\mathrm{d}k/\mathrm{d}\omega$
does not exist at $k=0$. Therefore, the temporal correlation function
$c\left(t\right)$ for these potentials is slowly decreasing. Nevertheless,
for the spatial correlation function, we have,
\begin{eqnarray}
h\left(k\right) & = & \int S\left(k,\omega\right)\mathrm{d}\omega\nonumber \\
 & = & 2V_{0}^{2}f\left(k\right).
\end{eqnarray}
If one chooses the function $f\left(k\right)$ to be infinitely differentiable
then following from the Riemann\textendash{}Lebesgue lemma the spatial
correlation function, $c\left(x\right)$ will decay faster than any
power law. A similar calculation could be repeated for the potentials
of the type (\ref{eq:2}), with the PSD (\ref{eq:Sk_optic}), leading
to a similar result. Therefore, a natural question to ask is whether
a necessary condition for the new universality class is a slowly decaying
correlation function. The answer to this question is negative as was
exemplified after (\ref{eq:Sk_essen_sing}).

In this subsection we have demonstrated that potentials of the type
(\ref{eq:1}) and (\ref{eq:2}), with any dispersion relation and
a fast decaying distribution of wave-numbers will result in a transport
slower than (\ref{eq:Dv_univ1}). Those potentials will have short
correlations in position and long correlations in time. For potentials
with smooth PSDs we have suggested a classification scheme based on
their PSD properties.

\section{Discussion\label{sec:Discussion}}

In the present work the diffusion coefficient for the velocity is
presented in terms of the spectral content, namely, the average power
spectral density (PSD), (see Eq. (\ref{eq:Dv_d})) for potentials
that are random both in space and time. This representation is very
natural for potentials generated as a superposition of waves used
in optics and atom optics. It is of particular use in the case where
the potential is a stationary process both in time and space. The
simplicity of the expression enabled to explore the properties of
the diffusion coefficient, to establish a scaling form of the Fokker-Planck
equation, and to discover new universality classes. In particular,
we were able to calculate explicitly the diffusion coefficient for
representative examples, relevant for applications both in optics
and atom optics (Gaussian and uniform distributions of wave-vectors).
We have shown that for dimensions larger than one only one universality
class is possible in the framework of the FP approximation, $D\left(v\right)\sim v^{-3}$.
However for one dimensional systems new possibilities for large velocity
asymptotics were also found.

In the limit of small velocities the diffusion coefficient vanishes
only when $S\left(k,0\right)\sim\delta\left(k\right)$, which is not
the typical behavior. However, for potentials described in Sec. \ref{sec:Potentials-sperposition}
$D\left(0\right)=0$ and moreover for small velocity the diffusion
coefficient grows faster than $v$, invalidating the Fokker-Planck
approximation for those potentials, in this regime.

\begin{figure}
\begin{centering}
\includegraphics{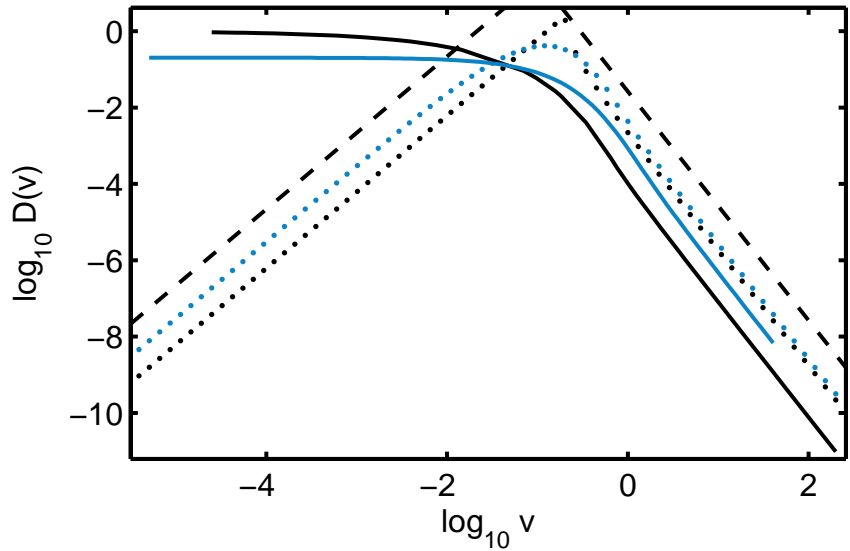}
\par\end{centering}

\caption{\label{fig:Dv_2d}Log-Log plot of the diffusion coefficients as a
function of the velocity for a two dimensional system and different
potentials. The dots stand for potentials of the form (\ref{eq:1}),
while the solid lines stand for potentials of the form (\ref{eq:2}).
The blue (light) line and dots are for Gaussian distribution of wave-numbers
while the black line and dots are for uniform distribution of wave-numbers
over a disc of unit radius. The dashed black lines are guides to the
eye with the slopes, $2$ and $-3$.}
\end{figure}

There is an important case were the waves composing the potential
satisfy a dispersion relation given by $\omega\left(k\right)\sim k^{2}$.
In this case we found that the diffusion coefficient depends on the
characteristic scale of the potential, $k_{R}$. In Fig. \ref{fig:Dv_2d}
$D\left(v\right)$ is presented for similar values of the characteristic
scale of the potential, $k_{R}$.

The main result of the paper is the classification of universality
classes in one dimensional systems presented in Sec. \ref{sec:Anomalous-diffusion-1D}.
In the past it was found that in the large velocity limit the diffusion
coefficient depends on the velocity as, $D\left(v\right)\sim v^{-3}$
\cite{Rosenbluth1992,Golubovic1991,Bezuglyy2006,Bezuglyy2012}. In
the present work we have shown that this is always the case for dimensions
larger than one, and explained the mechanism of this behavior (see
(\ref{eq:Dv_no_deltas})). However for one dimensional systems this
is only one of the possibilities. Generally it depends on the asymptotic
expansion of $D\left(v\right)$ in powers of $v^{-1}$:
\begin{enumerate}
\item The first term of the asymptotic expansion of $D\left(v\right)$ is
non-zero, $D\left(v\right)\sim v^{-3}$.
\item The first non-vanishing term in the asymptotic expansion of $D\left(v\right)$
is the $n-$th term then, $D\left(v\right)\sim v^{-n}$.
\item \emph{All} terms in the asymptotic expansion of $D\left(v\right)$
\emph{are} zero, $D\left(v\right)\leq v^{-\alpha}$, for any $\alpha>0$.
In particular, the diffusion coefficient may be zero for $v>v_{\max}$
(e.g. (\ref{eq:Dv_1d_segment})) or non-zero but decreasing faster
than any power law (e.g. (\ref{eq:Dv_1d_opt_Gauss})).
\end{enumerate}
All these possibilities can be realized in experiments with control
over the PSD or more precisely $f\left(k\right),$ as is clear from
(\ref{eq:Dv_cos_quad}) or (\ref{eq:Dv_opt_quad}). Unlike previous
studies \cite{Golubovic1991,Bezuglyy2006,Bezuglyy2012} the classification
into universality classes does not rely on the differentiability of
the correlation function of the potential. Additionally, as was demonstrated
in (\ref{eq:Cxt_essnt_sing}), the new universality classes does not
depend on the range of the correlation function. The general behavior
can be understood heuristically in terms of the Chirikov resonances
\cite{Zaslavskiii1972}.

We have also shown that initially particles will experience a uniform
acceleration and at some later stage, on a longer time-scale, the
velocity will follow diffusion as predicted by the Fokker-Planck equation.
Eventually, it will reach the asymptotic long-time behavior predicted
by the Fokker-Planck approximation.

In the present research the spreading of the velocity distribution
is studied in the framework of the Fokker-Planck equation, therefore
an obvious question to study is what happens when the Fokker-Planck
approximation fails. Another important issue, which was not addressed
in this work, is how to obtain analytically the spreading in position.
Here this spreading was studied numerically for the potentials (\ref{eq:1})
and (\ref{eq:2}). Since in the experiments that have motivated this
work, the relevant dynamics is of waves, rather than particles, an
obvious question to explore is the correspondence between the classical
and wave dynamics.

This work was motivated by the experimental work of Liad Levi and
Mordechai Segev, whom we thank for many stimulating discussions and
for providing crucial insight for this problem. It is our great pleasure
to thank Tom Spencer for introducing us to \cite{Golubovic1991,Rosenbluth1992,Bezuglyy2012}
and to Michael Wilkinson for introducing us to \cite{Bezuglyy2006,Bezuglyy2012}.
Many of the results of the present work originated from fruitful discussions
with Michael Wilkinson during his visit to the Technion. We would
also like to thank Igor Aleiner, Boris Altshuler and Michael Berry
for informative discussions. This work was partly supported by the
US-Israel Binational Science Foundation (BSF), by the Minerva Center
of Nonlinear Physics of Complex Systems, by the Shlomo Kaplansky academic
chair, by the Fund for promotion of research at the Technion.

\bibliographystyle{apsrev}
\bibliography{QP-potential}

\end{document}